\documentclass[prb,reprint,twocolumn,citeautoscript,amsmath,superscriptaddress]{revtex4-1}

\usepackage{bm,mathrsfs}
\usepackage{graphicx}
\usepackage{epsfig}
\usepackage{amsmath,bbm}
\usepackage{amsfonts,amssymb}
\usepackage{times}
\usepackage{ascmac}
\usepackage{here}
\usepackage{dsfont}
\usepackage{enumitem}
\usepackage[colorlinks,linkcolor=blue,citecolor=blue]{hyperref}
\usepackage{url}

\begin{document}

\title{Generative diffusion model with inverse renormalization group flows}
\author{Kanta Masuki}
\email{masuki@g.ecc.u-tokyo.ac.jp}
\affiliation{Department of Physics, The University of Tokyo, 7-3-1 Hongo, Bunkyo-ku, Tokyo 113-0033, Japan}
\author{Yuto Ashida}
\email{ashida@phys.s.u-tokyo.ac.jp}
\affiliation{Department of Physics, The University of Tokyo, 7-3-1 Hongo, Bunkyo-ku, Tokyo 113-0033, Japan}
\affiliation{Institute for Physics of Intelligence, The University of Tokyo, 7-3-1 Hongo, Tokyo 113-0033, Japan}

\begin{abstract}    
    Diffusion models represent a class of generative models that produce data by denoising a sample corrupted by white noise \cite{sohl-dickstein2015a,song2019,song2020a,ho2020b}. Despite the success of diffusion models in computer vision \cite{C2023},  audio synthesis \cite{kong2021audio}, and point cloud generation \cite{luo2021}, so far they overlook inherent multiscale structures in data and have a slow generation process due to many iteration steps. In physics,  the renormalization group offers a fundamental framework for linking different scales and giving an accurate coarse-grained model \cite{wilson1975,wilson1983}.
    Here we introduce a renormalization group-based diffusion model that leverages multiscale nature of data distributions for realizing a high-quality data generation. 
    In the spirit of renormalization group procedures, we define a flow equation that progressively erases data information from fine-scale details to coarse-grained structures. Through reversing the renormalization group flows, our model is able to  generate high-quality samples in a coarse-to-fine manner.
    We validate the versatility of the model through applications to protein structure prediction  \cite{A1973,D2008}  and image generation \cite{C2023}. Our model consistently outperforms conventional diffusion models across standard evaluation metrics, enhancing sample quality and/or accelerating sampling speed by an order of magnitude. The proposed method alleviates the need for data-dependent tuning of hyperparameters in the generative diffusion models, showing promise for systematically increasing sample efficiency based on the concept of the renormalization group.
\end{abstract}

\maketitle
\begin{figure*} 
    \centering
    \includegraphics[width=17.5cm]{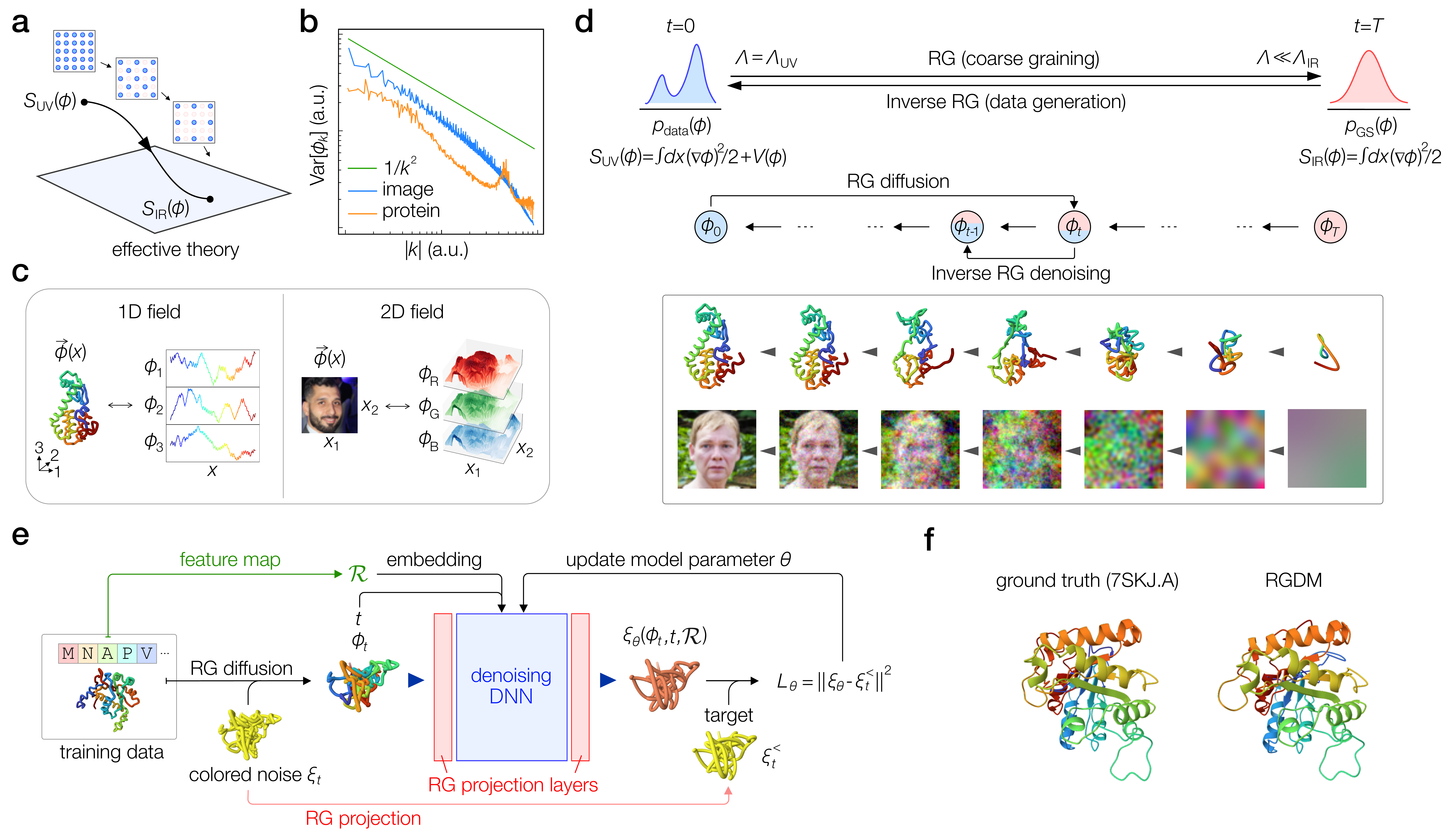}
    \caption{ 
        \label{fig1} {\bf Renormalization group-based diffusion model (RGDM), and its application to protein structure prediction and image generation.}
        {\bf a,} Schematic illustrating the concept of the RG theory. A complex microscopic model in the UV is coarse-grained into an effective IR model in a lower-dimensional subspace. 
            {\bf b,} Power law decay of the power spectral densities of natural data, which motivates the functional ansatz~\eqref{data_theory} for modeling a data distribution. 
        {\bf c,} Field representations of natural data. 
         A protein structure and image can be regarded as a vector field \(\vec{\phi}(x)\) on a one- and two-dimensional space, respectively.
        {\bf d,} Overview of the RGDM. Starting from a data distribution \(p_{\text{data}}\) at a short distance (or equivalently, a UV wavenumber $\Lambda_{\text{UV}}$), the RG process iteratively performs its coarse-graining, during which all the correlations up to a wavenumber scale $\Lambda$ are retained. As the RG scale $\Lambda$ is changed from $\Lambda_{\text{UV}}$ to a value below the IR scale $\Lambda_{\text{IR}}$, the model gradually flows to the Gaussian distribution. Through reversing these flows, the RGDM generates a sample in a coarse-to-fine manner; the insets show typical generation processes.  
        {\bf e,} RGDM architecture for training. The RGDM learns the colored noise \(\xi_t\) whose  schedule is judiciously fixed by the RG theory. To realize an efficient generation, we introduce the  projection layers before and after the denoising deep neural network (DNN), which remove the integrated-out high wavenumber modes.  In protein structure prediction, we embed the information of amino sequences \(\mathcal R\) into the DNN.
        {\bf f,} Typical protein structure generated by the RGDM compared with the ground truth. The total number of generation steps is \(T=98\).
    } 
\end{figure*}

The renormalization group (RG) is a framework that relates a microscopic model at short distances in the `ultraviolet' (UV) to its coarse-grained effective model at large scales in the `infrared' (IR)  \cite{wilson1975,wilson1983}, see Fig.~\ref{fig1}a.
In the last decades, the RG has proven to be a powerful tool to understand physics at radically different scales, such as the asymptotic freedom of elementary particles \cite{gross1973,politzer1973} and phase transitions in statistical and condensed matter physics \cite{kadanoff1966,fisher1974,polchinski1999,schollwock2005,bulla2008,metzner2012}. 
One of the key ingredients for the success of the RG is its multiscale nature, where systems that differ at short scales exhibit similar behaviour at coarse-grained scales. 
Among numerous ways for obtaining macroscopic descriptions of natural phenomena, the RG stands out because of its capability to rigorously eliminate fluctuations at short scales while preserving all the correlations at longer scales \cite{wetterich1993,polchinski1984a,morris1994}.

Multiscale structures are ubiquitous in nature and can be found in a wide range of data beyond physical sciences \cite{ruderman1994a,vanderschaaf1996a,saremi2013a}. Consider, for example, a face image. Due to correlations among nearby pixels, one can probably recognize what is depicted in a blurred image that retains the main characteristics while losing the fine details. 
More specifically, the power spectral densities of various data, such as natural images and protein structures, universally exhibit an approximate power law $\propto k^{-2}$ with wavenumber $k$ in a similar manner as typically observed in physics (Fig.~\ref{fig1}b). This fact motivates us to explore a way to leverage inherent multiscale properties in data for realizing their efficient generation based on the RG. 

To link the RG with generative models, we start from identifying a local variable in each sample by a `field' variable (Fig.~\ref{fig1}c), making natural data amenable to RG procedures. 
The evolution of a coarse-grained distribution when changing the length scale of interest is then governed by the RG flow equation akin to a convection-diffusion process \cite{cotler2023b}. 
Our key observation here is that a similar type of flow equations have been discussed in diffusion models, a class of generative models that produce data by denoising a sample corrupted by white noise \cite{song2019,song2020a,ho2020b}. 
Generally, diffusion models first define a forward process that progressively destroy data by adding white noise to each data dimension, gradually converting a data distribution \(p_{\text{data}}\) to the uncorrelated Gaussian distribution\cite{sohl-dickstein2015a}.
These models are then trained to learn the denoising function that removes the added noise. Thereafter, samples are generated by running the step-by-step reverse process, where the models start from white noise and iteratively refine it to produce a realistic data sample.

Despite the success of diffusion models, they have so far completely overlooked multiscale structures in data, and their slow generation process currently remains as a crucial drawback; to obtain a high-quality sample, each generation typically requires  hundreds or thousands of iteration steps \cite{C2023,yang2023a}.  
The inevitable tradeoff between quality and computational cost has been a bottleneck in expanding the applicability of iterative generative models to important domains, such as protein structure prediction \cite{A1973,D2008}.
While AlphaFold has revolutionalized protein structure prediction from sequence information \cite{jumper2021}, it tends to fail to predict dynamical aspects of proteins, such as conformations of intrinsically disordered proteins \cite{chakravarty2022,tesei2024,chakravarty2024}. Hence, there is a growing interest in a generative  model that produces a distribution of protein structures rather than a single most probable configuration \cite{jing2023,nakata2023,qiao2024}. 

In this paper, we introduce a renormalization group-based diffusion model (RGDM) that generates data through reversing the RG flows (Fig.~\ref{fig1}d). The forward process in our model implements the coarse-graining of the RG by adding judiciously chosen colored noises, where local information is progressively erased from fine-scale details to coarse-grained structures. More specifically, we employ the exact RG \cite{polchinski1984a,kopietz2010}, a nonperturbative framework for rigorously defining the wavenumber-space procedure of integrating out short-distance modes, to define a flow equation that systematically contracts a complex data distribution into a coarse-grained distribution in a lower-dimensional subspace (Fig.~\ref{fig1}a). Following the reverse flows, the model generates a sample in a coarse-to-fine manner by leveraging the multiscale structures in data.    
This feature makes a contrast with existing diffusion models such as denoising diffusion probabilistic models (DDPMs) \cite{ho2020b}, which are based on white noise and thus overlook the multiscale nature.

One of the main characteristics of the RGDM is its low computational cost during both training and generation. While the DDPMs simultaneously denoise all the wavenumber modes at every iteration step, the scale separation between low- and high-wavenumber modes in the RGDM naturally allows us to reduce this redundancy (Fig.~\ref{fig1}e), making the training stable and accelerating the sampling while preserving sample quality. 
We validate the efficiency and versatility of the RGDM through applications to real-world problems in two domains, namely, protein structure prediction \cite{A1973,D2008}  and image generation \cite{C2023}; see Fig.~\ref{fig1}f for a typical example in the former. 
We demonstrate that our model consistently outperforms the DDPMs across standard datasets in terms of the sample quality and/or sampling speed. For instance, in image generation, the RGDM requires only hundreds of steps to generate a sample image whose quality is comparable to the one produced by a thousand-step DDPM. 
We find that reducing the number of iteration steps has little impact on the sample quality in the RGDM, indicating that the tradeoff between quality and computational cost can be  significantly alleviated. 
Finally, we discuss how our model can serve as a guiding principle in choosing hyperparameters of diffusion models. 
In the RGDM, all the noise schedules are unambiguously determined by the RG theory once a regulator is specified. 
Thus, our model has much less ambiguity than DDPMs, the latter of which require heuristic tuning of hyperparameters depending on specific properties, such as image size or protein length  \cite{kingma2021,chen2023,hoogeboom2023,jing2023}. 
\\
\\
\noindent
\textbf{Renormalization group approach to natural data}
\\
To illustrate how the idea of the RG can be applied to natural data, we consider a microscopic model $S_{\text{UV}}(\phi)$ at a short-distance scale (or equivalently, a UV scale), which gives a probability functional $p(\phi)\propto e^{-S_{\text{UV}}(\phi)}$ for a sample $\phi$. We assume that the model has the shortest distance scale $a$,  which, for instance, corresponds to the size of each pixel in image or the distance between two neighboring $\alpha$-carbons in protein. In the wavenumber space, this scale defines the largest wavenumber $\Lambda_{\text{UV}}=2\pi/a$  called a UV cutoff. Suppose that we have access only to long-distance scales corresponding to wavenumber scales less than $\Lambda$. It is then natural to introduce a coarse-grained effective description by eliminating higher-wavenumber modes as \cite{wilson1983}  
\begin{align}
    S_{ \text{eff},\Lambda}(\phi^<) &= -\ln \int [d\phi^>] e^{-S_{\text{UV}}(\phi^<+\phi^>)}.\label{RG_concept}
\end{align}
Here, \(\phi^<\) (\(\phi^>\)) denotes the Fourier modes with wavenumber below (above) \(\Lambda\). 
The RG gives a way to rigorously describe how an effective model~\eqref{RG_concept} changes as \(\Lambda\) is gradually decreased.

\begin{figure*} 
    \centering
    \includegraphics[width=17.5cm]{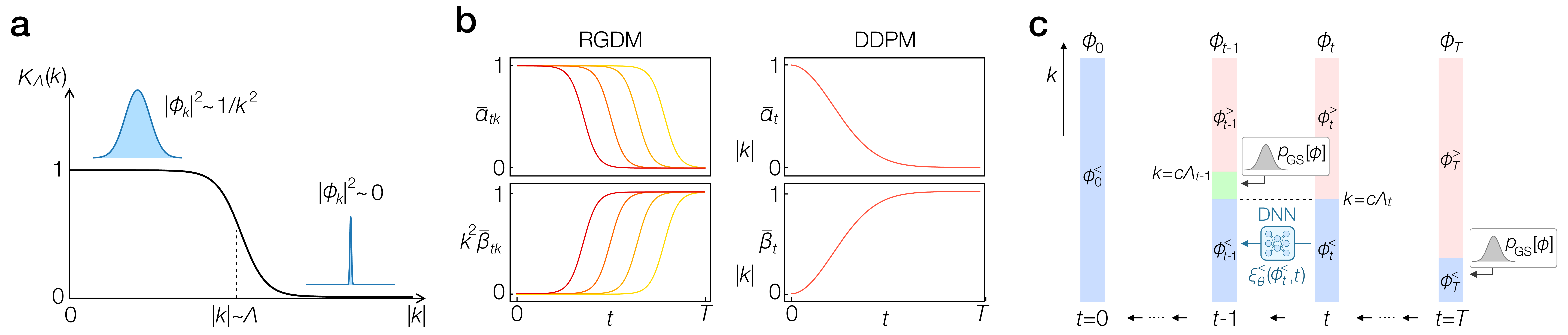}
    \caption{\label{fig2} 
    {\bf Cutoff function, noise schedule, and sampling scheme.}
    {\bf a,} Cutoff function $K_{\Lambda}(k)$ in the exact RG. The value of $K_\Lambda(k)$ continuously alters from one to zero around \(|k|\sim\Lambda\),  which allows for eliminating higher-wavenumber fluctuations, ensuring the scale-separation property of the RG. 
    {\bf b,} Noise parameters plotted as a function of the logarithmic RG scale $t$. The noise schedule in the RGDM (left panels) is unambiguously determined from the exact RG theory in such a way that the forward process progressively destroys Fourier modes from high- to low-wavenumber components by adding colored noise to a sample. The scaling $k^2\bar{\beta}_{tk}\simeq 1$ at large $t$ ensures that the model eventually converges to the fixed-point Gaussian distribution \(\phi\!\sim\!p_{\text{GS}}\propto\exp(-\frac{1}{2}\int_k k^2|\phi_k|^2)\). The DDPMs (right panels) use the white noise, whose schedule is the same for all the Fourier modes and needs to be heuristically fine-tuned in a data-dependent manner. 
    {\bf c,} Sampling scheme by the inverse RG flows from $t=T$ to $0$. Due to the scale-separation property in Eq.~\eqref{scale_sep}, the higher wavenumber components \(\phi_{t}^>\) are eliminated from the model and obey the Gaussian distribution \(p_{\text{GS}}\).  
    In the generation process at a step \(t\), one creates  \(\phi_{t-1}\) by performing the denoising only on the lower-wavenumber modes \(\phi_t^<\) (blue color). The Fourier modes in the wavenumber shell (green color) are the newly integrated-out components and sampled from the Gaussian distribution \(p_{\text{GS}}\).  The modes $\phi_{t-1}^>$ (red color) are not sampled at all as they have been already eliminated from the model.}
\end{figure*}

As noted in the introduction, fluctuations in natural data often exhibit characteristic spectral properties, which we can use to infer a plausible functional ansatz for an effective model. 
In particular, the $k^{-2}$ decay observed in the variance of Fourier modes \(\phi_k\) in Fig.~\ref{fig1}b implies the presence of the scale-invariant quadratic contribution \(\int_x(\nabla\phi)^2/2\), leading to an ansatz,
\begin{align}
    S_{\text{data}}(\phi) = \frac{1}{2}\int_x\ (\nabla\phi(x))^2 + V(\phi)\label{data_theory},
\end{align}
which gives a data distribution $p_{\text{data}}(\phi)\propto e^{-S_{\text{data}}(\phi)}$. 
Here, \(\int_x\equiv\int d^dx\) denotes the integral over the $d$-dimensional coordinate space, and \(V(\phi)\) includes  higher-order contributions.
The standard RG procedure can be readily applied to natural data because a theory in the form of Eq.~\eqref{data_theory} is nothing but a model typically encountered in statistical physics. Starting from $p_{\text{data}}$ as a microscopic model at a UV scale, one can systematically perform its coarse-graining by utilizing the multiscale structures therein. 

To be concrete, we use the exact RG \cite{polchinski1984a,kopietz2010} to obtain an effective model \(S_\Lambda(\phi)\) whose distribution \(p_\Lambda(\phi)\!\propto\!e^{-S_\Lambda(\phi)}\) explicitly satisfies the scale-separation property
\begin{align}
    p_\Lambda(\phi) = p_{\text{eff},\Lambda}(\phi^<) p_{\text{GS}}(\phi^>),\;\;p_{\text{GS}}(\phi) \!\propto\! e^{-\frac{1}{2}\int_x(\nabla\phi)^2},\label{scale_sep}
\end{align}
where  \(p_{\text{GS}}\) is the Gaussian (GS) distribution that has no information about the data distribution \(p_{\text{data}}\), while \(p_{\text{eff},\Lambda}\) is an effective distribution that retains all the correlations in data up to the wavenumber scale \(\Lambda\). As detailed below, the evolution of $p_{\Lambda}$ with $\Lambda$ is governed by a certain convex-diffusion equation called the flow equation, and the corresponding one-parameter family, $p_{\Lambda}$, constitutes the RG flows. 

It is important to note that, in practice, any realistic data have the largest length scale $L$, which, for instance, corresponds to the size of an image or the length of a protein. This scale naturally defines a IR cutoff \(\Lambda_{\text{IR}}\sim 1/L\) in the wavenumber space. 
When $\Lambda$ is tuned far below $\Lambda_{\text{IR}}$, all modes will be essentially eliminated. Thus, starting from any data distribution \(p_{\text{data}}\) at a UV scale, its effective distribution $p_{\Lambda}$ ultimately flows to the Gaussian distribution \(p_{\text{GS}}\) in the IR limit $\Lambda\!\to\!0$.
This property allows our generative model to make sampling by reversing the RG flows starting from the known distribution \(p_{\text{GS}}\).
\\

\noindent
\textbf{Renormalization group-based diffusion model}
\\
We introduce an RG-based generative model, where each sample is obtained by starting from the Gaussian sampling \(\phi\sim p_{\text{GS}}\) in the deep IR regime $\Lambda\ll\Lambda_{\text{IR}}$ and then performing the step-by-step inverse RG procedures up to the UV scale $\Lambda=\Lambda_{\text{UV}}$ (Fig.~\ref{fig1}d). To this end, we start from defining a forward process that implements the coarse-graining of the exact RG. Let us write an effective model as \cite{kopietz2010}
\begin{align}
    S_\Lambda(\phi) = \frac{1}{2}\int_k K^{-1}_\Lambda(k)k^2 |\phi_k|^2 + V_{\Lambda}(\phi),\label{Pol_SL}
\end{align}
where \(\int_k\equiv\int\frac{d^dk}{(2\pi)^d}\) denotes the integral over the wavenumber space, and  \(K_\Lambda(k)\) is a cutoff function whose value continuously changes from one to zero around \(|k|\!\sim\!\Lambda\) as \(|k|\) is increased, see Fig.~\ref{fig2}{a}. 
This choice of \(K_\Lambda\) implements the scale-separation property such that the higher-wavenumber modes approximately obey the delta distribution $\phi^>\sim\delta(|\phi^>|)$, while the statistics of the other modes \(\phi^<\) are characterized by an effective model \(S_{\text{eff},\Lambda}(\phi^<)\equiv S_\Lambda(\phi)|_{\phi^>=0}\). 
It is customary to represent  \(K_\Lambda\) by 
\begin{align}
    K_\Lambda(k) = \frac{r(k^2/\Lambda^2)}{1+r(k^2/\Lambda^2)},\label{KL}
\end{align}
where $r(x)$ is a monotonically decreasing function called regulator. 
We note that the only ambiguity in the exact RG lies in a choice of the regulator \(r(x)\), whose optimal choice has been the issue in theoretical physics~\cite{litim2000a,litim2001a}.
We find that a simple regulator $r(x)=x^{-1}$, a common choice in physics, suffices to generate high-quality image data, while another choice can be useful for protein structure prediction (see Methods).

To ensure that the flows of $S_\Lambda$ faithfully realize the RG procedures, all the correlations must be preserved up to a wavenumber scale $\Lambda$ during the flows. 
Polchinski has shown that this condition can be satisfied by renormalizing \(V_\Lambda\) in Eq.~\eqref{Pol_SL} in accordance with the elimination of the fluctuations around \(|k|\sim\Lambda\) as \cite{polchinski1984a} (see Supplementary Information): 
\begin{align}
    \partial_\Lambda V_{\Lambda} = -\frac{1}{2}&\int_k\ k^{-2}\partial_\Lambda K_\Lambda(k)\left(\frac{\delta^2 V_{\Lambda}}{\delta\phi_k\delta\phi_{-k}}
    - \frac{\delta V_{\Lambda}}{\delta\phi_k}\frac{\delta V_{\Lambda}}{\delta\phi_{-k}} \right).\label{Pol_flow}
\end{align}
While one may define a forward process according to Eq.~\eqref{Pol_flow}, we find that the resulting generative model in this case fails to generate a high-quality sample due to inefficient training in the IR regime, where the singularity arises from the delta distribution $\delta(|\phi|)$. 
To avoid this difficulty and realize stable training of the RGDM, we find it useful to rescale the distribution as $p_\Lambda(\phi) \propto e^{-S_\Lambda(\sqrt{K_\Lambda}\phi)}$.   
In this way, the Gaussian distribution \(p_{\text{GS}}\) in Eq.~\eqref{scale_sep}, which underlies the ansatz~\eqref{data_theory}, becomes the steady-state solution of the flow equation (or equivalently, the fixed point of the RG flows).
Accordingly, the effective distribution \(p_\Lambda\) no longer has the singularity described above, making training and sampling stable. 

Taken together, the forward process of the RGDM can be given by the  convex-diffusion equation\cite{cotler2023b} (see Supplementary Information)
 \begin{align}
    \partial_t p_t(\phi) = -\frac{1}{2}\int_k\left[ \frac{\partial_tK_t}{k^2 K_t}\frac{\delta^2p_t}{\delta\phi_k\delta\phi_{-k}}+ \frac{\delta}{\delta\phi_k}\left(\frac{\partial_tK_t}{K_t}\phi_kp_t\right)\right],\label{rgdm_FP}
\end{align}
where `time' $t\equiv \tau\ln(\Lambda_{\text{UV}}/\Lambda_t)$ denotes a logarithmic RG scale with a time constant $\tau$, giving  \(p_t\equiv p_{\Lambda_t}\) and  \(K_t\equiv K_{\Lambda_t}\). One can realize the RG forward diffusion~\eqref{rgdm_FP} by adding wavenumber-dependent colored noises to each sample,  
\begin{align}
    \phi_{tk} &= \sqrt{\bar\alpha_{tk}}\phi_{0k} + \sqrt{\bar\beta_{tk}}\epsilon_k,\label{rgdm_fdiff}\\
    \bar\alpha_{tk} &= K_{t}(k),\ \bar\beta_{tk} = k^{-2}(1-\bar\alpha_{tk}), 
\end{align}
where $\epsilon_k\sim {\cal N}(0,1)$ is the standard normal noise, and the initial condition at  $t=0$ is determined by sampling from the data distribution at a UV scale, i.e., $\phi_{t=0}\sim p_{\text{data}}$ at $\Lambda_{t=0}=\Lambda_{\text{UV}}$. 
The added noises eliminate fluctuations around $|k|\sim\Lambda_t$ and thereby destroy fine-scale details first and then progress toward coarse-grained structures. 
We again emphasize that the time and wavenumber dependence of the noise schedule (cf. \(\bar\alpha_{tk}\) and \(\bar\beta_{tk}\) in Eq.~\eqref{rgdm_fdiff}) is unambiguously determined by the RG theory once the one-dimensional function $r(x)$ in Eq.~\eqref{KL} is specified, see Fig.~\ref{fig2}{b}. 
This makes a sharp contrast to existing diffusion models, where a noise schedule needs to be fine-tuned without solid theoretical backgrounds \cite{C2023,yang2023a}.

The generation in the RGDM can be now implemented by reversing the RG flows. Specifically, we discretize the time at \(t\in\{0,1,\ldots, T\}\) and let the denoising DNN learn the colored noise $\xi_{\theta k}$ that approximates $\sqrt{\bar\beta_{tk}}\epsilon_k$ in Eq.~\eqref{rgdm_fdiff}, see Methods and Fig.~\ref{fig1}e. The generation process starts from sampling \(\phi_T\!\sim\! p_{\text{GS}}\) by the Gaussian distribution and then iteratively performs its denoising from \(t\!=\!T\) to \(0\), where each sample is created in a coarse-to-fine manner (Fig.~\ref{fig2}c). 
To leverage the scale-separation property~\eqref{scale_sep} for reducing the computational cost, we add the projection layers before and after the DNN (Fig.~\ref{fig1}e). These layers remove redundant Fourier modes with \(|k|\!>\!\Lambda_t\), which have been already eliminated by the coarse-graining procedures of the RG. 
This simplification reduces the effective dimensions of the input and output vectors of the denoising DNN, enabling both the stable training and the efficient generation of high-quality samples. 
\\

\begin{figure*}[t] 
    \centering
    \includegraphics[width=17.5cm]{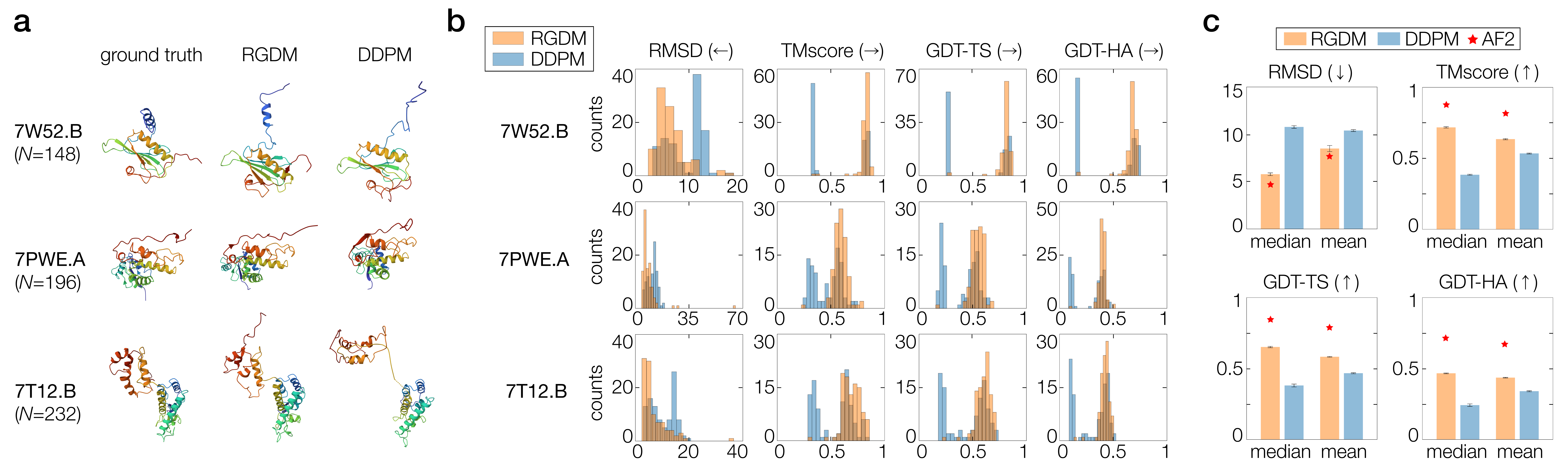}
    \caption{\label{fig3}
    {\bf Protein structure prediction by the RGDM.}
    {\bf a,} Typical protein structures generated by the RGDM and DDPM. We randomly choose proteins from CAMEO targets, which are not included in the training dataset, and predict the protein structure once for each model. The total number of generation steps \(T\) is a function of the amino-sequence length \(N\) and set to be \(T\!=\!T_{0}\!+\!\tau\ln(N/N_0)\) so that the RG scale at the final diffusion step scales as \(\Lambda_T\propto N^{-1}\). We use the parameters \(T_{0}=80, \tau= 8.17, N_0=32\) in both models.
    {\bf b,} Comparisons of the sample quality between the RGDM and DDPM. We assess the quality based on standard evaluation metrics and plot the histograms by generating samples 100 times for each of the proteins in Fig.~\ref{fig3}{a}. We note that a lower RMSD indicates the better sample quality, while larger values mean the higher-quality samples for the other metrics. 
    {\bf c,} Single-structure prediction accuracy evaluated on the CAMEO targets composed of 182 proteins. We make the plots by generating a protein structure once for each target and calculating the medians and means of the metrics across the sampled structures. The red stars indicate the reference values attained by AlphaFold2 (AF2). }
\end{figure*}

\noindent
\textbf{Applications to protein structure prediction and image generation}
\\
To demonstrate the versatility of the RGDM, we first apply it to protein structure prediction. 
Our attempt is complementary to other state-of-the-art architectures such as AlphaFold, which deterministically predicts a single most probable static structure \(\phi(\mathcal R)\) from the amino sequence represented by a feature tensor \(\mathcal R\) \cite{jumper2021,wu2022}. 
We here, in contrast, aim to use the RGDM to stochastically generate protein structures from the \(\mathcal R\)-conditioned distribution \(p_{\mathcal R}(\phi)\). 
We train the model on the Protein Data Bank \cite{berman2000protein} (PDB) and test it by the Continuous Automated Model EvaluatiOn (CAMEO) targets. For all datasets, we focus on predicting a structure of the \(\alpha\)-carbon chain in proteins and benchmark our model by making comparison with the DDPM, whose noise schedule and other hyperparameters are chosen according to the known heuristics \cite{ho2020b}. 

Figure~\ref{fig3}{a} shows typical protein structures generated by the RGDM and DDPM with the total number of iteration steps $T\simeq90$. To evaluate the sample quality, we use several standard metrics that allow one to measure the structural similarity of proteins, including the root-mean-square deviation (RMSD), TM-score, the global distance test-total score and -high accuracy (GDT-TS and GDT-HA, respectively). We generate samples 100 times for each of the proteins in Fig.~\ref{fig3}{a} and plot the histograms of the metrics in Fig.~\ref{fig3}{b}. We find that the RGDM accurately and precisely generates high-quality samples in terms of the above metrics, as evidenced by the sharp single peaks in the histograms. In contrast, the results obtained by the DDPM exhibit the bimodal distributions and fail to accurately predict protein structures. We also show the single-structure prediction accuracy evaluated on the CAMEO targets in Fig.~\ref{fig3}{c}, where a protein structure is sampled once for each target. 
While there still exist discrepancies from state-of-the-art performance by the AlphaFold2 (red stars),   
the results clearly demonstrate that the RGDM can outperform the DDPM across the evaluation metrics.

\begin{figure*}
    \centering
    \includegraphics[width=17.5cm]{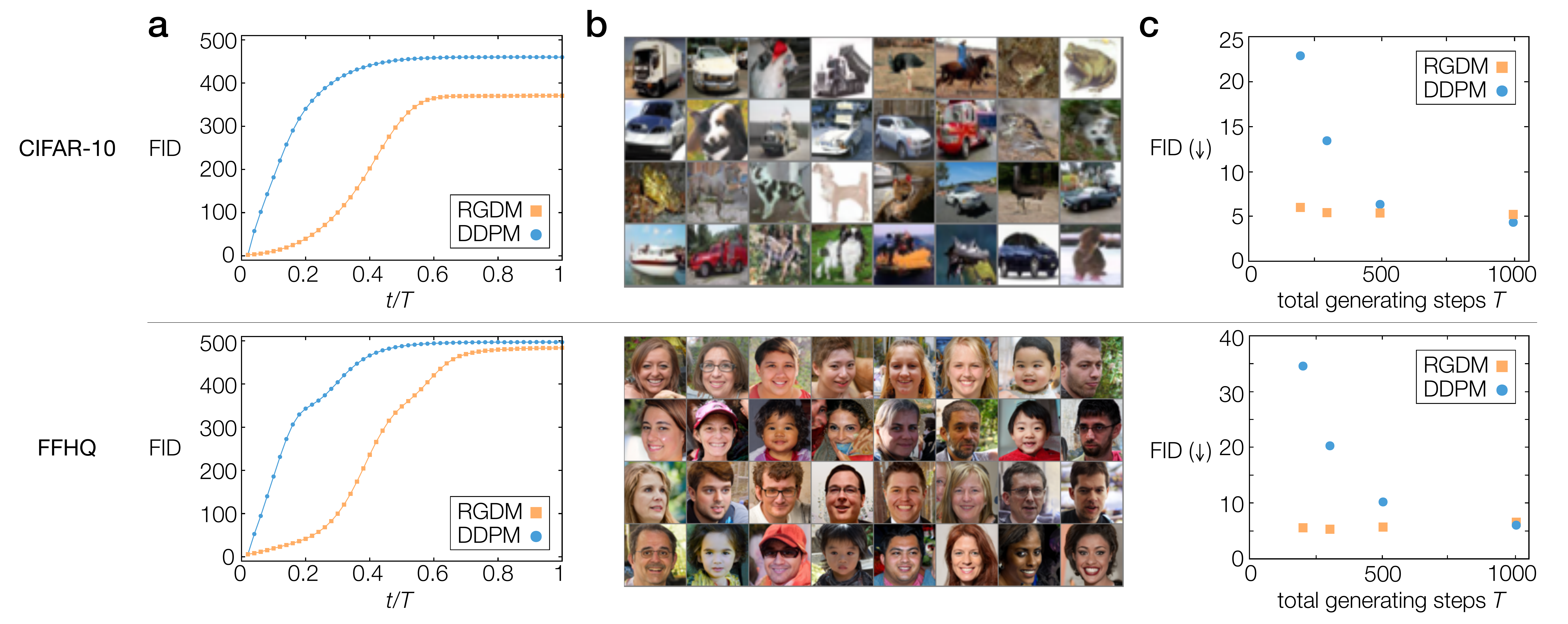}
    \caption{\label{fig4}
    {\bf Image generation by the RGDM.} Top (bottom) panels show the results for the CIFAR-10 (FFHQ) dataset. We use CIFAR-10 images with resolution \(32\times 32\) and FFHQ images resized to \(64\times 64\).
    {\bf a,} Distance between the data distribution \(p_{\text{data}}\) and an intermediate distribution \(p_t\) in the forward diffusion measured  by the frech\'et interception distance (FID) \cite{heusel2017a, parmar2021}. We take the total number of generation steps to be \(T=1000\).
    {\bf b,} Typical images generated by the RGDM that are unconditionally trained on the CIFAR-10 and FFHQ datasets with \(T=200\).
    {\bf c,} Sampling quality of the RGDM and DDPM trained on the CIFAR-10 and FFHQ datasets at different total generation steps \(T\). Each data point is obtained by evaluating the FID between the training and sampled datasets obtained by generating \(5\times 10^4\) image samples. An error bar is negligibly small compared to the marker size in the plots.
   }
\end{figure*}

We next benchmark our model against the DDPM on image generation tasks with different resolutions, namely, CIFAR10 (32$\times$32) \cite{cifar10} and FFHQ (64$\times$64) \cite{ffhq}. 
Figure~\ref{fig4}{a} plots the frech\'et interception distance (FID) \cite{heusel2017a,parmar2021}, which measures the distance between the data distribution \(p_{\text{data}}\) and an intermediate distribution \(p_t\) during the forward diffusion. In both datasets, we find that \(p_t\) in the DDPM rapidly moves away from \(p_{\text{data}}\) at small \(t\), indicating that most of the meaningful data structures are created in the last few steps of the generation process. In contrast, the RGDM gradually breaks the initial distribution \(p_{\text{data}}\) over a long period of time in the forward diffusion, which allows for generating fine structures with high accuracy. Typical images generated by the RGDM are shown in Fig.~\ref{fig4}{b}.

To make a further quantitative comparison between the RGDM and DDPM, we evaluate the sampling quality of each model trained with different generation steps by the FID between the training dataset and sampled images (Fig.~\ref{fig4}{c}). We here note that the RGDM  defines the noise schedule regardless of image resolution and generation steps \(T\), while the DDPM uses a data-dependent heuristic tuning of the schedule (see Supplementary Information). 
In both CIFAR10 and FFHQ datasets, we find that the RGDM achieves higher sample quality than the DDPM, where the superiority becomes more evident as the total number of steps $T$ is decreased. While the DDPM is able to generate high-quality samples when $T$ is an order of thousand, the RGDM can achieve comparable or even better quality with much fewer steps. 
The RGDM can thus accelerate the sampling at a minor cost in sample quality. 
These results suggest that the tradeoff between quality and computational cost can be  alleviated by employing the RG framework for data generation. 
\\

\noindent
\textbf{Discussions}
\\
We leveraged the rigorous RG framework to develop a new class of generative models that produce each sample in a coarse-to-fine manner by reversing the RG flows. 
Through applications to protein structure prediction and image generation, we demonstrated that our model can consistently outperform the existing diffusion model regardless of size and dimension of data, improving the sample quality and/or accelerating the sampling speed.  
The proposed model can be applied to a wide range of data without tuning the noise schedule, which will allow for harnessing multiscale structures to systematically enhance sample efficiency.

The present results are complementary to several ongoing efforts in the field of generative models. For instance, our method should be combined with a variety of the known heuristics \cite{song2020a,kong2021,kingma2021,karras2022a,ho2022a,saharia2023a,chen2023,hoogeboom2023} for making sampling of the DDPM efficient, which could accelerate the data generation and thus widen the applicability of the generative diffusion models. Meanwhile, our RG-based approach provides a general framework for diffusion models constructed in the wavenumber space,  giving a unified view for comparing different types of proposed noise schedules \cite{hoogeboom2022a,rissanen2022,ho2022a,saharia2023a} (see Supplementary Information).  
It is noteworthy that the arbitrariness in the RGDM remains in a choice of the regulator \(r(x)\) in Eq.~\eqref{KL}; the above studies might provide insights into an optimal choice of  \(r(x)\) toward improving the performance of the RGDM.

From a technical perspective, a refinement of the denoising DNN could improve the prediction ability of our model. In the present calculations, we added the projection layers before and after the DNN, but the structure of the DNN itself could be also optimized to fully harness the scale-separation property of the RG. Such a network optimization might further reduce the computational cost in the training and sampling. 
Finally, we note that many state-of-the-art generative models, including  AlphaFold3 \cite{abramson2024}, incorporate a diffusion model as a submodule. While so far most of them are based on conventional diffusion models such as the DDPM, replacing them by the RGDM might give a simple way to enhance sample efficiency and/or quality of those architectures. 
\\
\\
\noindent
{\bf Acknowledgements}\\
We are grateful to Yoshiyuki Kabashima, Kyogo Kawaguchi, Hosho Katsura, Masato Koashi, Shunshi Kohyama, Ziyin Liu, Jun Mochida, Ryuna Nagayama, Kazumasa A. Takeuchi, Hidenori Tanaka, and Masahito Ueda for fruitful discussions. 
K.M. acknowledges support from the Japan Society for the Promotion of Science (JSPS) through Grant No.~JP24KJ0898. Y.A. acknowledges support from JSPS through Grant No. JP19K23424 and from JST FOREST Program (Grant Number JPMJFR222U, Japan) and JST CREST (Grant Number JPMJCR23I2, Japan). 

\bibliographystyle{naturemag_arxiv}

\break
\noindent
{\bf Methods}
\\
{\bf Training and generation in the RGDM}\\
The training of the RGDM is implemented by discretizing the forward process at \(t\in\{0,1,\ldots, T\}\) and letting the denoising DNN learn the colored noise with the cost function,
\begin{align}
    L_\theta = \sum_{t=1}^T \lambda_t\ \mathbb E_{\phi_0\sim p_{\text{data}},\xi_t\sim\mathcal{N}(0,\bar\beta_t)}
    \left[||\xi_\theta(\phi_t,t)-\xi_t||^2\right],\label{cost}
\end{align}
where the noise vector $\xi_t$ is related to a sample vector by \(\phi_{tk} = \sqrt{\bar\alpha_{tk}}\phi_{0k}\!+\!\xi_{tk}\) (cf.  Eq.~\eqref{rgdm_fdiff}). We set the wavenumber scale at the final forward step to be \(\Lambda_T\!\ll\!\Lambda_{\text{IR}}\) so that the final distribution in the diffusion process, \(p_{T}\), approximately becomes the Gaussian distribution \(p_{\text{GS}}\).
We find it useful to choose \(\lambda_t\) in such a way that \(\lambda_t\propto (\sum'_k \bar\beta_{tk})^{-1}\) (\(\lambda_t\propto (\sum'_k 1)^{-1}\)) in image generation (protein structure prediction), where \(\sum'_k\) denotes the summation over the wavenumbers after the RG projection (see Supplementary Information).
 
We generate data by firstly sampling \(\phi_T\!\sim\! p_{\text{GS}}\) and performing the denoising process from \(t=T\) to \(0\) as 
\begin{align}
    \phi_{t-1k} &= \frac{1}{\sqrt{\alpha_{tk}}}\left(\phi_{tk}-\frac{\beta_{tk}}{\bar\beta_{tk}}\xi_{\theta k}(\phi_t, t)\right) + \sqrt{\beta_{tk}}\epsilon_k,\label{rgdm_bdiff}\\
    \alpha_{tk} &= \bar\alpha_{tk}/\bar\alpha_{t-1k},\ \ 
    \beta_{tk} = k^{-2}(1-\alpha_{tk}).
\end{align}
As discussed in the main text, we simplify the training and the sampling of the RGDM based on the scale-separation property of the RG in Eq.~\eqref{scale_sep}. Specifically, at the \(t\)-th step of the diffusion, Eq.~\eqref{scale_sep} implies that the higher wavenumber components \(\phi_{k>\Lambda_t}\) reduce to the Gaussian noise \(\xi_{tk}=\sqrt{\bar\beta_{tk}}\epsilon_k\), which has no information about the data.  
Thus, the only noise that should be learned by the denoising DNN is \(\xi^<(\phi_t^<, t)\), i.e., the lower-wavenumber components with $|k|<\Lambda_t$ (cf. Fig.~\ref{fig2}{c}).
We include the projection layers before and after the DNN to remove the redundant Fourier components with \(|k|\!>\!\Lambda_t\), thereby changing the learning objective from \(\xi(\phi_t,t)\) to \(\xi^<(\phi_t^<, t)\) (Fig.~\ref{fig1}{e}). 
This replacement effectively reduces the dimensions of the input vector (from \(\phi_t\) to \(\phi^<_t\)) and the output vector (from \(\xi_t\) to \(\xi^<_t\)) of the DNN, making the training of the RGDM stable and efficient.
\\

\noindent
{\bf Protein structure prediction}
\\
In the applications to the protein structure prediction in this study, we trained the RGDM and DDPM on all the protein structures deposited on the PDB before Apr 30, 2020 (232646 proteins in total). In evaluating the validation error, we used the validation dataset composed of all the protein structures deposited between May 1, 2020 and Nov 30, 2020 (14201 proteins in total). In the assessment of the single-structure prediction accuracy in Fig.~\ref{fig3}{b}, we used CAMEO dataset composed of all the CAMEO targets released between Aug 1, 2022 and Oct 31, 2022.
We used the features \(\mathcal R\) generated by OmegaFold \cite{wu2022}, which utilizes a language model in creating amino-sequence features \(\mathcal{R}\), and excluded the CAMEO targets with 750 or more residues, for which OmegaFold cannot generate a feature tensor. The number of the resulting CAMEO targets are 182 in total.

We trained the e3NN~\cite{thomas2018,e3nn,jing2023} to learn the colored noise as a function of \(\phi_t, t,\) and \(\mathcal R\) (cf. Fig.~\ref{fig1}{e}). 
By using the backward diffusion~\eqref{rgdm_bdiff} with fixed \(\mathcal R\), we generated the protein structures \(\phi\) that obey the $\cal R$-conditioned distribution \(p_{\mathcal R}(\phi)\). We found it useful to choose the regulator function \(r(x)=(\exp(\ln^2(x+1)-1))^{-1}\) (see Supplementary Information). 
To make a fair comparison between the RGDM and DDPM, we used the same denoising DNN and the same \(T\) in both models. The noise schedule of the DDPM is determined according to the known heuristics \cite{ho2020b} (see Supplementary Information), while the RGDM uses the colored noise whose schedule is derived from the RG theory as explained in the main text. 
We trained the RGDM and DDPM with two different choices of generation steps \(T\), leading to 4 trained models in total. The number of total generation steps \(T\) for each protein with sequence length \(N\) was chosen to be $T = T_0 + \tau \ln(N/N_0)$ with constants $\tau=8.17, T_0=80,$ and $N_0=32$. We trained each model using the protein training dataset describe above with 10 training epochs. It takes about two weeks with a single NVIDIA GeForce. 
During the training, we updated the model parameters by Adam optimizer with the exponential moving average (EMA) ratio 0.9999. 
After the training, we chose the best epoch based on the sample quality evaluated with the validation dataset described above. We used these models to make Fig.~\ref{fig3}.

In Fig.~\ref{fig3}{c}, we sampled a protein structure once for each CAMEO target (182 proteins in total) and calculated the median and mean of the quality metrics. We repeated this procedure 10 times and estimated the error (see Supplementary Information). We note that the RGDM typically fails to sample protein structures at a rate of approximately once every 15 attempts. When the model fails to generate a protein sample, we assigned the value of 0 to the TM-score, GDT-HA, and GDT-TS, while RMSD was set to be infinity. The mean RMSD was then evaluated on the samples excluding the ones leading to the divergent results. During the sampling by AlphaFold2, we generated each structure once, performing one recycling step without any atom relaxation.
\\

\noindent
{\bf Image generation}
\\
In the image generation, we used the FFHQ dataset (thumbnails 128\(\times\)128) \cite{ffhq} and the CIFAR-10 training dataset. To reduce the training time, we resized the FFHQ images with resolution \(128\times128\) to \(64\times64\), while we used the original \(32\times32\) images in the CIFAR-10 dataset. To avoid poor antialiasing reported in Ref.~\cite{parmar2021}, we resized FFHQ images by PIL library with bicubic interpolation. The FFHQ dataset (CIFAR-10 dataset) includes $7\times 10^4$ ($5\times 10^4$) images. We used the regulator function \(r(x)=x^{-1}\) and trained each of the models, RGDM and DDPM, with varied total generation steps \(T\). On the CIFAR-10 dataset, we trained a model at \(T=200, 300, 500, 1000\) with $2\times 10^6$ training steps, which takes about 72 hours for each model with a NVIDA RTX6000Ada. On the FFHQ dataset, we trained a model at \(T=200, 300, 500, 1000\) with $10^6$ training steps, which takes about one week for each model with the same machine.
During the training, we updated the model parameters by Adam optimzer with the EMA ratio 0.9999, and we saved the model parameters every $5\times 10^4$ steps. After the training, we evaluated the FID for each checkpoint by generating \(10^4\) samples. We then used the best checkpoint and reevaluated the FID by generating \(5\times 10^4\) samples to create Fig.~\ref{fig4}.
We used the same UNet \cite{ho2020b} as the denoising DNN in both RGDM and DDPM (see Supplementary Information).
\\

\noindent
{\bf Power law decay of spectral densities in natural data}\\
To obtain the plot in Fig.~\ref{fig1}{b}, we used the data taken from the PDB and the FFHQ dataset described above. For each sample, we calculated the Fourier components \(\phi_{k}\) by performing the orthonormal discrete cosine transform (DCT). We then calculated the variance of \(\phi_{k}\) as 
\begin{align}
    {\text{Var}}[\phi_{k}] = \mathbb E\left[||\phi_{k}-\bar\phi_{k}||^2\right],\label{Meq:var}
\end{align}
where $\mathbb E[\cdot]$ represents the ensemble average, and \(\bar\phi_{k}=\mathbb E [\phi_k]\) denotes an expectation value.  
We note that the DCT wavevector \(k\) takes \(k\!=\!i\pi/N\) (\(0\leq\!i\!\leq N\!-\!1\)) for proteins with \(N\) amino acids and \(k\!=\!(i\pi/L,j\pi/L)\) (\(0\!\leq\!i,j\leq L\!-\!1\)) for image data with resolution \(L\times L\).
Since proteins with different length \(N\) have different DCT wavevector \(k\), one cannot directly take the average~\eqref{Meq:var}. Instead, we used proteins with length \(N=32,50,90,150,230\)  and calculated the variance~\eqref{Meq:var} for each \(N\), where 400\(\sim\)1300 proteins in the protein dataset are included for each \(N\). Finally, we merged the variance and created the plot of the protein data in Fig.~\ref{fig1}{b}. 
In making the plot of the image data in Fig.~\ref{fig1}{b}, we eliminated the directional dependency of the variance~\cite{vanderschaaf1996a,rissanen2022}. Specifically, we introduced a small wavenumber \(\Delta k\) and used the averaged variance instead of Eq.~\eqref{Meq:var} as 
\begin{align}
    \underset{|k|<|k'|<|k|+\Delta k}{\text{average}}{\text{Var}}[\phi_{k'}].
\end{align}
We used 5000 randomly chosen image data from the FFHQ dataset.
\\

\noindent
{\bf Data availability}\\
Data that support the plots in this paper are available on GitHub: \url{https://github.com/kantamasuki/RGDM}. The image datasets used in this paper were taken from \url{https://www.cs.toronto.edu/~kriz/cifar.html} and \url{https://github.com/NVlabs/ffhq-dataset}. The protein structural data used in this study were provided by the Protein Data Bank. 
\\

\noindent
{\bf Code availability}\\
Code, all the training parameters and trained models used to generate the results in this paper can be found at \url{https://github.com/kantamasuki/RGDM}.

\widetext
\pagebreak
\begin{center}
\textbf{\large Supplementary Materials}
\end{center}

\renewcommand{\theequation}{S\arabic{equation}}
\renewcommand{\thefigure}{S\arabic{figure}}
\setcounter{equation}{0}
\setcounter{figure}{0}

\subsection{\label{Ss:RG}Overview of the renormalization group}
We here give an overview of the renormalization group (RG) and explain the details that are necessary to understand the key idea behind our work. 
To make the contents accessible to a broad range of research communities, we present this section in a self-contained manner. 
In Sec.~\ref{Sss:Effec}, we first illustrate the concept of the RG and an effective model. 
In Sec.~\ref{Sss:Dim_an} we explain the dimensional analysis that allows one to infer the relevance of each term in effective models in the long-distance limit. 
Based on the scaling dimension, we also discuss the reason why the model \(S_{\rm data}(\phi)\) in the form of Eq.~(2) in the main text can be considered as a meaningful ansatz to describe natural data.
In Secs.~\ref{Sss:Func} and~\ref{Sss:Polc}, we introduce the functional representation of correlation functions and derive the Polchinski RG equation (Eq.~(6) in the main text). 
The derivation of the corresponding diffusion equation (Eq.~(7) in the main text) is provided in Sec.~\ref{Sss:diff}.

\subsubsection{\label{Sss:Effec}Effective model}
\begin{figure}[b]
    \centering
    \includegraphics[width=5cm]{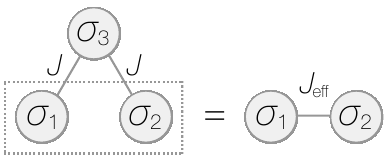}
    \caption{\label{Sfig:three_spins}The effective model of the first two spins can be obtained by decimating the third spin from the original model. In the effective model, there arises an effective interaction \(J_{\rm eff}\) between \(\sigma_1\) and \(\sigma_2\), which is mediated by the third spin in the original model.}
\end{figure}

In the RG, one aims to obtain an effective model at a certain length scale by decimating shorter-distance fluctuations that are out of interest \cite{wilson1975,*wilson1983}.
We shall first illustrate the notion of such a decimation with a toy example. Let us consider the probability distribution \(p(\bm \sigma)\) of a three-spin system defined as 
\begin{align}
    p(\bm \sigma) = \frac{e^{-H(\bm \sigma)}}{Z},\ H(\bm \sigma) = -J(\sigma_1\sigma_3 + \sigma_2\sigma_3)\label{Seq:H_spin},
\end{align}
where each spin \(\sigma_i\) takes \(\pm1\), and \(Z=\sum_{\bm \sigma} e^{-H(\bm \sigma)}\) is the normalization constant to ensure \(\sum_{\bm \sigma}p(\bm \sigma)=1\), which is known as the partition function in statistical physics (see Fig.~\ref{Sfig:three_spins}).
 Suppose that we have access only to the first two spins \((\sigma_1,\sigma_2)\).  It is then natural to consider the following distribution, which decimates the third spin \(\sigma_3\) as
\begin{align}
    p_{\rm eff}(\sigma_1,\sigma_2) 
    &\equiv \sum_{\sigma_3=\pm 1} p(\sigma_1,\sigma_2,\sigma_3)=\frac{e^{-H_{\rm eff}(\sigma_1,\sigma_2)}}{Z_{\rm eff}},\label{Seq:peff_spin}\\
    H_{\rm eff}(\sigma_1,\sigma_2) &= -J_{\rm eff}\sigma_1\sigma_2,\\
    J_{\rm eff} &= \frac{1}{2}\ln\frac{e^{2J}+e^{-2J}}{2}.
\end{align}
Importantly, as long as one focuses on the first two spins, the model \(H_{\rm eff}(\sigma_1,\sigma_2)\) gives the same predictions with the original model \(H(\bm \sigma)\). A model constructed in such a way is called an \textit{effective model}.
We note that, while the direct interaction between \(\sigma_1\) and \(\sigma_2\) is absent in the original model \(H(\bm \sigma)\), the decimation of the third spin gives rise to an interaction between \(\sigma_1\) and \(\sigma_2\) in the effective model \(H_{\rm eff}(\sigma_1,\sigma_2)\). In this way, a decimation of unnecessary degrees of freedom in general {\it renormalizes} interactions between the remaining degrees of freedom, making the effective model highly nontrivial.

The decimation procedure described above can be readily extended to field theory on the basis of  the path-integral formalism \cite{Negele98,kopietz2010}.
In field theory, one introduces a field \(\phi\), which is a scalar or vector-valued function on the Euclidian space \(\mathbb R^d\). A field theory is then defined by a functional \(S[\phi]\) of the field \(\phi\), which is called action or simply model. While we use the round-bracket notation \(S(\phi)\) in the main text, here we shall use the square-bracket notation \(S[\phi]\), following a convention in field theory to emphasize that \(S\) is a functional. 
The action \(S[\phi]\) provides a probability functional of $\phi$ as follows: 
\begin{align}
    p[\phi] = \frac{e^{-S[\phi]}}{Z},\ Z = \int[d\phi]\; e^{-S[\phi]},\label{Seq:p_phi}
\end{align}
where \(\int[d\phi] (\cdots)\) denotes the functional integral called the path integral, which is the integration over all possible configurations of the field \(\phi\) \cite{FH10}. 
In fact, long-distance physics of a wide class of classical/quantum many-body systems can be described by a field theory by appropriately defining the field \(\phi\) and the model \(S[\phi]\) \cite{Hubbard59,Negele98}.

In the field-theoretical description, the information about the system can be obtained through the correlation functions 
\begin{align}
    \langle \phi_{\alpha_1}\cdots\phi_{\alpha_n}\rangle_S \equiv \int[d\phi] \frac{e^{-S[\phi]}}{Z}\phi_{\alpha_1}\cdots\phi_{\alpha_n},\label{Seq:corr_def}
\end{align}
where \(\alpha_i\) is the argument of the field \(\phi\) such as the position \(x\) or the wavenumber \(k\). 
Therefore, based on Eq.~\eqref{Seq:corr_def}, one constructs an effective model so that the correlations in the original model \(S[\phi]\) for relevant degrees of freedom are preserved.
For example, if we are interested in the macroscopic behaviour of the model \(S[\phi]\) above the length scale \(\Lambda^{-1}\), we decompose the field \(\phi\) into the small- and large-wavenumber fluctuations, \(\phi^<\) and \(\phi^>\), based on the wavenumber scale \(\Lambda\) as
\begin{align}
    \phi &= \phi^< + \phi^>,\\
    \phi^>_k &= \theta(k^2-\Lambda^2)\phi_k,\\
    \phi^<_k &= \theta(\Lambda^2-k^2) \phi_k.\label{Seq:field_decomp}
\end{align}
Here, $\phi_k$'s are the Fourier modes of the field, and \(\theta(x)\) is the Heaviside step function.
We then consider the effective model \(S_{\rm eff,\Lambda}[\phi^<]\) defined as (cf. Eq.~\eqref{Seq:peff_spin})
\begin{align}
        e^{-S_{{\rm eff},\Lambda}[\phi^<]} &= \int[d\phi_>] e^{-S[\phi^<+\phi^>]}.\label{Seq:eff_theory}
\end{align}
We note that, by definition, the effective model \(S_{{\rm eff},\Lambda}[\phi^<]\) preserves all the correlations below the cutoff scale as
\begin{align}
    \langle \phi_{k_1}\cdots \phi_{k_n}\rangle_{S_{{\rm eff},\Lambda}} 
    &= \langle \phi_{k_1}\cdots \phi_{k_n}\rangle_{S}\ \ \text{for } |k_i|<\Lambda,\label{Seq:corr_eff_theory}
\end{align}
where \(\langle \cdot \rangle_{S_{{\rm eff},\Lambda}}\) denotes the expectation value with respect to  the effective model \(S_{\rm eff}[\phi]\). For the sake of notational simplicity, below we express the effective model \(S_{\rm eff,\Lambda}[\phi]\) by \(S_\Lambda[\phi]\).

\subsubsection{\label{Sss:Dim_an}Dimensional analysis}
Although the formal expression of the effective model \(S_{\Lambda}[\phi]\)~\eqref{Seq:eff_theory} is exact, it is usually very difficult to directly obtain such a model. 
Instead, one often employs a perturbative RG approach \cite{wilson1983}, where the path-integral is approximately performed by a perturbative expansion of interaction terms while neglecting irrelevant terms based on the scaling dimension \cite{Gol1992,Nishi2011}. 
More generally, It is the dimensional analysis that motivates such a treatment; in this section, we  thus give a summary of the dimensional analysis and explain the notion of relevance in the effective field theory.
We note that the discussion here intuitively explains the reason why \(S_{\rm data}[\phi]\) (Eq.~(2) in the main text) can be regarded as a plausible ansatz for natural data distributions.

To illustrate the dimensional analysis, we start with the effective model \(S_\Lambda[\phi]\) at RG scale in Eq.~\(\Lambda\)~\eqref{Seq:eff_theory}, whose field \(\phi\) is expressed in the Fourier space as
\begin{align}
    \phi(x) &= \int_{|k|<\Lambda} e^{ikx} \phi_k.\label{Seq:phi_x}
\end{align}
Here, we denote the integral \(\int \frac{d^dk}{(2\pi)^d}\) by \(\int_k\). 
To examine the evolution of the effective model, let us slightly decrease the cutoff scale \(\Lambda\) to \(\Lambda'\!=\!\Lambda/b\) (\(b\!>\!1\)) and compare \(S_\Lambda\) with \(S_{\Lambda'}\). To this end, we first decompose the field \(\phi\) based on the new cutoff \(\Lambda'\) as 
\begin{align}
    \phi^<(x) &= \int_{|k|<\Lambda'} e^{ikx} \phi_k,\ \ 
    \phi^>(x) = \int_{\Lambda'\leq|k|<\Lambda} e^{ikx} \phi_k.
\end{align}
Also, we decompose \(S_\Lambda[\phi]\) into three parts: the ones that depend only on \(\phi^<\) or \(\phi^>\), respectively, and the one that depends on both \(\phi^<\) and \(\phi^>\),
\begin{align}
    S_\Lambda[\phi] &= S_\Lambda^<[\phi^<] + S_\Lambda^>[\phi^>] + S_\Lambda^{\rm int}[\phi^<,\phi^>].
\end{align} 
The effective model at the new RG scale \(\Lambda'\) is then expressed as
\begin{align}
    S_{\Lambda'}[\phi^<] &= S_{\Lambda}^<[\phi^<] 
    - \ln\left\langle e^{-S_\Lambda^{\rm int}[\phi^<,\phi^>]}\right\rangle _{S_\Lambda^>[\phi^>]} + {\rm const.},\label{Seq:da_effgrow}
\end{align}
where we use Eq.~\eqref{Seq:eff_theory}, and \(\langle \cdot \rangle_{S_\Lambda^>[\phi^>]}\) is the expectation value with respect to the model \(S_{\Lambda}^>[\phi^>]\).
We note that \(S_{\Lambda'}[\phi^<]\) in Eq.~\eqref{Seq:da_effgrow} cannot be still directly compared with \(S_\Lambda[\phi]\) since they have different UV cutoffs \(\Lambda'\) and \(\Lambda\). Therefore, we rescale the field \(\phi^<(x)\) as 
\begin{align}
    \tilde\phi(x) = b^{[\phi]} \phi^<(bx),\label{Seq:da_res}
\end{align}
where we introduce a constant \([\phi]\) called the \textit{scaling dimension} of \(\phi\). Since \(S_{\Lambda'}\) as a functional of \(\tilde\phi\) has a UV cutoff at \(\Lambda\), it can be now directly compared with \(S_{\Lambda}[\phi]\).
In particular, one can examine the RG `flow' based on the infinitesimal change of \(S_{\Lambda'}[\tilde\phi]\) when continuously increasing the value of \(b\geq 1\).
In fact,  the flow of the contribution  \(S_\Lambda^<[\phi^<]\) in \(S_{\Lambda'}[\tilde\phi]\)  in Eq.~\eqref{Seq:da_effgrow} can be simply determined by the dimensional analysis, providing a qualitative picture of the RG flow as we discuss now.

To be concrete, we consider an effective model \(S_{\Lambda}[\phi]\) in the form
\begin{align}
    S_{\Lambda}[\phi] &= c_\Lambda \int_x (\nabla\phi(x))^2 + m_\Lambda\int_{x} \phi^2(x) + u_\Lambda\int_x \phi^4(x),\label{Seq:pert1}
\end{align}
where \(\phi\) is the field with the cutoff wavenumber \(\Lambda\) as in Eq.~\eqref{Seq:phi_x}. Thus, \(S_\Lambda^<[\phi^<]\) in Eq.~\eqref{Seq:da_effgrow} becomes
\begin{align}
    S^<_{\Lambda}[\phi^<] &= c_\Lambda \int_x (\nabla\phi^<(x))^2 + m_\Lambda\int_{x} (\phi^{<}(x))^2 + u_\Lambda\int_x (\phi^{<}(x))^4\nonumber\\
    &= b^{[c]}c_\Lambda\int_x(\nabla\tilde\phi(x))^2 + b^{[m]} m_\Lambda\int_x\tilde\phi^2(x) + b^{[u]} u_\Lambda \int_x \tilde\phi(x)^4\label{Seq:pert2}
\end{align}
where we rewrite \(S^<_\Lambda[\phi^<]\) as a functional of \(\tilde\phi(x)\) by using  Eq.~\eqref{Seq:da_res}. 
Here, \([c], [m]\), and \([u]\) are the scaling dimension of each coefficient, \(c_\Lambda, m_\Lambda\), and \(u_\Lambda\), given by \([c] = d-2-2[\phi]\), \([m]=d-2[\phi]\), and \([u]=d-4[\phi]\), respectively.
Therefore, \(S_{\Lambda'}[\tilde\phi]\) can be written as 
\begin{align}
    S_{\Lambda'}[\tilde\phi] &= c_{\Lambda'} \int_x (\nabla\phi(x))^2 + m_{\Lambda'}\int_{x} \phi^2(x) + u_{\Lambda'}\int_x \phi^4(x) + \cdots,\\
    c_{\Lambda'} &= b^{[c]} c_\Lambda + \cdots,\label{Seq:cl}\\
    m_{\Lambda'} &= b^{[m]} m_\Lambda + \cdots,\label{Seq:ml}\\
    u_{\Lambda'} &= b^{[u]} u_\Lambda + \cdots,\label{Seq:ul}
\end{align}
where the residual terms expressed by \((\cdots)\) denote the nontrivial contributions from the second term in the right hand side of Eq.~\eqref{Seq:da_effgrow}.
Due to the contributions in Eqs.~\eqref{Seq:cl}-\eqref{Seq:ul}, the evolution of each coefficient when increasing \(b\) has a contribution determined by the scaling dimension as 
\begin{align}
    \frac{dc_l}{dl} &= [c] c_l + \cdots,\\
    \frac{dm_l}{dl} &= [m] m_l + \cdots,\\
    \frac{du_l}{dl} &= [u] u_l + \cdots.\\
\end{align}
Here, we introduce the logarithmic RG scale \(l=-\ln\Lambda\), which is also denoted by `time' $t$ in the main text, and represent the coefficients as a function of \(l\).
With these expressions, it is important to note that any coefficient with a large negative scaling dimension would quickly decay when \(l\) is increased (i.e., \(\Lambda\) is decreased), thereby having little effects on the effective model in the IR limit $\Lambda_0$. Conversely, coefficients with positive scaling dimension can rapidly grow when \(l\) is increased and are expected to play important roles in the effective model.
In this way, one can qualitatively evaluate the relevance of each term in the long-distance behaviour based on the scaling dimension.

So far, the scaling dimension of each coefficient depends on the scaling dimension of the field \([\phi]\), which is introduced in Eq.~\eqref{Seq:da_res}. 
In fact, to determine \([\phi]\), it is necessary to first choose which term we use as a reference to infer  the relevance of the other terms \cite{polchinski1999}.
For example, suppose that one examines the RG flow of the effective model~\eqref{Seq:pert1} by using the quadratic kinetic term, \(\int_x(\nabla\phi)^2\), as a reference. In this case, one defines the value of \([\phi]\) in such a way that the scaling dimension of \(c\), the coefficient of \(\int_x(\nabla\phi)^2\), becomes zero. Accordingly, \([\phi]\) is set to be \((d-2)/2\), which results in \([m]=2\) and \([u]=4-d\). 
In other words, \(\int_x \phi^2(x)\) is always relevant while \(\int_x \phi^4(x)\) becomes irrelevant when \(d>4\). 
In particular, if the relevant term \(\int_x\phi^2(x)\) is present in the UV regime, it grows during the course of the RG, and \(\int_x(\nabla\phi)^2\) alone will no longer faithfully describe the long-distance behaviour of the model. For another example, if we evaluate the relevance of the other term by choosing \(\int_x \phi^2(x)\) as a reference,  we need to take \([m]=0\), which results in \([\phi]=d/2\), \([c]=-2\), and \([u]=-d\). In this case, the other terms are always irrelevant regardless of the dimension \(d\). 
More generally, the expression of the scaling dimension for a term including $n$ derivatives and $m$ fields is given by
\begin{align}
    g_\Lambda \int d^dx\ \left( 
        \underbrace{\partial\cdots\partial}_{n \text{ derivatives in total}}\ 
        \underbrace{\phi(x)\cdots\phi(x)}_{m \text{ fields in total}}\right)
        \ \ \to \ \ 
        [g] = d-n-m[\phi].\label{Seq:gL}
\end{align}

We now discuss why the model \(S_{\rm data}(\phi)\) in Eq.~(2) in the main text can be regarded as a suitable ansatz for natural data distributions. 
To this end, we first note that the quadratic terms, \(\int d^dx\ \partial^m\phi\partial^n\phi\)\ (\(m,n=0,1,\ldots\)), can be written down in order of relevance as \(\int_x \phi^2(x)\), \(\int_x(\partial\phi(x))^2\), and so on, where we drop \(\int_x\phi(x)\partial\phi(x)\) since it is a integral of the total derivative \(\partial(\phi^2(x)/2)\). 
We note that all the terms in the effective model become irrelevant when compared to \(\int_x\phi^2(x)\) since it gives the scaling dimension of \(\phi\) as \([\phi]=d/2\) (cf. Eq.~\eqref{Seq:gL}). Therefore, if \(\int_x\phi^2(x)\) is present in the UV model, the other terms are expected to quickly decay along the RG flow. In other words, the coarse-grained model rapidly converges to the Gaussian model characterized by \(\int_x\phi^2(x)\), which simply corresponds to the white noise. Empirically, we know that this is not the case for natural data; even if the data, such as images of human faces, are coarse-grained to some extent, they still retain the characteristics of the original data, allowing us to recognize them as faces. It is thus feasible to argue that \(\int_x\phi^2(x)\) is absent (or at least has a negligibly small coefficient) in the UV model corresponding to a natural data distribution. In contrast, we do not find any compelling reason to prohibit the contribution \(\int_x(\partial\phi(x))^2\). In fact, this term favors minimizing \(|\phi(x)-\phi(x+dx)|^2\), which is consistent with what is commonly observed in natural data; for example, nearby pixels in an image basically tend to have similar colors, and nearest neighbor amino acids in a protein are typically positioned close to each other in real space. Thus, the model \(S_{\rm data}(\phi)\) in Eq. (2) in the main text, which has \(\int_x(\partial\phi(x))^2\) as the quadratic term, can be regarded as a plausible ansatz for the model of natural data distributions.

\subsubsection{\label{Sss:Func}Correlation functions in field theory}
While the perturbative RG with the dimensional analysis is a powerful tool to examine the long-distance properties of various models with weak interaction strengths, the  exact RG gives a nonperturbative framework for implementing the RG in the model at all coupling strengths \cite{polchinski1984a,wetterich1993,bonini1993,morris1994,kopietz2010}. For instance, while the perturbative RG often fails to track the RG flow correctly when the UV model has an irrelevant term (in the sense of dimensional analysis) with a large coupling, the framework of the exact RG or the functional RG \cite{kopietz2010} gives a systematic way to examine such systems by incorporating nonperturbative effects \cite{KY2021,TKY2023}.
In this section, we introduce functional representations of correlation functions; this formulation is necessary to derive the exact RG flow equation known as the Polchinski equation in the next section.

To be concrete, we consider a model \(S[\phi]\) with an \(\mathbb R\)-valued vector field \(\phi(x) = (\phi_1(x),\ldots,\phi_M(x))\) on \(\mathbb R^d\) as
\begin{align}
    S[\phi] &= S_0[\phi] + V[\phi],\label{Seq:S}\\
    S_0[\phi] &= \frac{1}{2}\int_\alpha\int_{\alpha'} \phi_\alpha [G_{0}^{-1}]_{\alpha\alpha}\phi_{\alpha'}.\label{Seq:S0}
\end{align}
Here, \(S_0[\phi]\) is the quadratic part, and \(V[\phi]\) is the higher-order interaction part,  \(\alpha = (x,i)\) denotes a set of the indices of the field \(\phi(x)=(\phi_i(x))_{i=1,\ldots,M}\), and the symbol \(\int_\alpha\) denotes the integration and the summation with respect to \(x\) and \(i\). 
We assume that the Gaussian propagator \(G_0\) is isotropic and has a translational symmetry as
\begin{align}
    [G_0]_{xi,yj} = \delta_{ij} G_0(x-y),
\end{align}
which simplifies Eq.~\eqref{Seq:S0} to
\begin{align}
    S_0[\phi] &= \frac{1}{2} \int \frac{d^dk}{(2\pi)^d} G^{-1}_0(k)|\phi_k|^2,
\end{align}
where we define the Fourier transform of \(G_0\) as \(G_0(k)=\int d^dx\ e^{-ikx}G_0(x)\).
For the sake of notational simplicity, it is convenient to regard the field \(\phi_\alpha = \phi_i(x)\) as a ``vector'' with indices \(\alpha=(x,i)\); below, we use the inner-product-like notation such as 
\begin{align}
    S_0[\phi] = \frac{1}{2}(\phi, G_0^{-1}\phi).
\end{align}

As describe above, all the information about the model \(S[\phi]\) can be obtained through the correlation functions (cf. Eq.~\eqref{Seq:corr_def})
\begin{align}
    G^{(n)}_{\alpha_1\cdots\alpha_n} 
    \equiv \int [d\phi]\ \frac{e^{-S[\phi]}}{Z}\ \phi_{\alpha_1}\cdots\phi_{\alpha_n}\ \ (n=1,2,\ldots),\label{Seq:corr_funcs}
\end{align}
where \(Z\) is the partition function of \(S[\phi]\) defined as \(Z =\int [d\phi] e^{-S[\phi]}\). The correlation functions \(G^{(n)}_{\alpha_1\cdots\alpha_n}\) have the following representation in terms of the generating functional \(\mathcal G[J]\) as
\begin{align}
    \mathcal G[J] & \equiv \frac{1}{Z}\int [d\phi]\ e^{-S[\phi] + (J,\phi)},\\
    G_{\alpha_1\cdots\alpha_n}^{(n)} & = \frac{\delta^n \mathcal{G}[J]}{\delta J_{\alpha_1}\cdots\delta J_{\alpha_n}}\biggl|_{J=0}.\label{Seq:G_func_rep}
\end{align}
Here, we introduce the external field \(J\) and the functional derivative \(\delta/\delta J_\alpha\) define by \(\delta J_{\alpha}/\delta J_{\alpha'} = \delta_{\alpha\alpha'} =\delta(x-x')\delta_{ii'}\). Indeed, the derivative of \(\mathcal{G}[J]\) reads 
\begin{align}
    \frac{\delta^n \mathcal{G}[J]}{\delta J_{\alpha_1}\cdots\delta J_{\alpha_n}} 
    &= \frac{\delta^{n-1}}{\delta J_{\alpha_1}\cdots\delta J_{\alpha_{n-1}}} \frac{1}{Z} \int[d\phi]\ \left(\frac{\delta}{\delta J_{\alpha_n}}(J,\phi)\right) e^{-S[\phi]+(J,\phi)}\\
    &= \frac{\delta^{n-1}}{\delta J_{\alpha_1}\cdots\delta J_{\alpha_{n-1}}} \frac{1}{Z} \int[d\phi]\ \phi_{\alpha_n} e^{-S[\phi]+(J,\phi)}\\
    &= \frac{1}{Z} \int [d\phi]\ \phi_{\alpha_1}\cdots\phi_{\alpha_n} e^{-S[\phi]+(J,\phi)},
\end{align}
where we use \(\frac{\delta}{\delta J_\alpha}(J,\phi)=\frac{\delta}{\delta J_\alpha}\int_\beta J_\beta\phi_\beta =\int_\beta \delta_{\alpha\beta}\phi_\beta = \phi_\alpha\).

In fact, the correlation functions \(G_{\alpha_1\cdots\alpha_n}^{(n)}\) can further be decomposed into more fundamental components.
To see this, we take the logarithmic of \(\mathcal{G}[J]\) and introduce the connected Green functions \(G_{{\rm c},\alpha_1\cdots\alpha_n}^{(n)}\) as  
\begin{align}
    \mathcal G_{\rm c}[J] &\equiv \ln\left(\frac{Z}{Z_0}\right) + \ln \mathcal G[J],\label{Seq:Gc_def}\\
    G_{{\rm c},\alpha_1\cdots\alpha_n}^{(n)} &\equiv \frac{\delta^n \mathcal G_{\rm c}[J]}{\delta J_{\alpha_1}\cdots\delta J_{\alpha_n}}\biggl|_{J=0}.\label{Seq:Gc_rep}
\end{align}
Here, \(Z_0\) is the partition function of the quadratic model \(S_0[\phi]\) defined as \(Z_0=\int[d\phi]e^{-S_0[\phi]}\). 
The correlation functions \(G_{\alpha_1\cdots\alpha_n}^{(n)}\) can be decomposed into the connected Green functions \(G^{(n)}_{{\rm c},\alpha_1\cdots\alpha_n}\) by taking all the possible contractions of \(\{\alpha_1,\ldots,\alpha_n\}\). For example, we have
\begin{align}
    G^{(1)}_{\alpha_1} &= G^{(1)}_{{\rm c},\alpha_1},\label{Seq:corr_decomp1}\\
    G^{(2)}_{\alpha_1\alpha_2} &= G^{(2)}_{{\rm c},\alpha_1\alpha_2} + G^{(1)}_{{\rm c},\alpha_1}G^{(1)}_{{\rm c},\alpha_2},\label{Seq:corr_decomp2}\\
    G^{(3)}_{\alpha_1\alpha_2\alpha_3} &= G^{(3)}_{{\rm c},\alpha_1\alpha_2\alpha_3} + G^{(2)}_{{\rm c},\alpha_1\alpha_2}G^{(1)}_{{\rm c},\alpha_3} + G^{(2)}_{{\rm c},\alpha_1\alpha_3}G^{(1)}_{{\rm c},\alpha_2} + G^{(2)}_{{\rm c},\alpha_2\alpha_3}G^{(1)}_{{\rm c},\alpha_1} + G^{(1)}_{{\rm c},\alpha_1}G^{(1)}_{{\rm c},\alpha_2}G^{(1)}_{{\rm c},\alpha_3}.\label{Seq:corr_decomp3}
\end{align}
We note that these decompositions can be proved by sequentially applying the derivative \(\frac{\delta}{\delta J_{\alpha_1}}\cdots\frac{\delta}{\delta J_{\alpha_n}}\) to both sides of \(\mathcal G[J]\!=\!e^{\mathcal G_c[J]+{\rm const.}}\), using the formula \((\delta/\delta J_\alpha)F_1[J]F_2[J] = (\delta F_1[J]/\delta J_\alpha) F_2[J] + F_1[J](\delta F_2[J]/\delta J_\alpha)\).
We mention that the term ``connected'' refers to the fact that the connected Green functions can be obtained by means of diagrammatic expansion with connected Feynman diagrams \cite{Negele98}.

To examine the property of connected Green functions, it is convenient to rewrite \(\mathcal G_{\rm c}[J]\) in Eq.~\eqref{Seq:Gc_def} as 
\begin{align}
    \mathcal G_c[J] 
    &= \ln\left(
        \frac{1}{Z_0} \int [d\phi] e^{-S[\phi] + (J,\phi)}
        \right)\\
    &= \ln\left(
        \frac{1}{Z_0} \int [d\phi]\ e^{\frac{1}{2}(J,G_0J)}e^{-S_0[\phi-G_0J]-V[\phi]}
    \right)\\
    &= \frac{1}{2}(J,G_0J) + \ln \left(
        \int[d\phi] \frac{e^{-S_0[\phi]}}{Z_0} e^{-V[\phi + G_0J]}
    \right)\\
    &= \frac{1}{2}(J, G_0J) + \ln \langle e^{-V[\phi+G_0J]} \rangle_{S_0[\phi]},\label{Seq:Gc_gen_formal}
\end{align}
where we change the integration variable from \(\phi\) to \(\phi+G_0J\) in the second line of the equations.
Also, \(\langle \cdot\rangle_{S_0[\phi]}\) is the expectation value with respect to the quadratic model \(S_0[\phi]\). With Eq.~\eqref{Seq:Gc_gen_formal}, we obtain the formal expression of \(\mathcal G_c[J]\) as 
\begin{align}
    G_{\rm c}[J] = \frac{1}{2}(J,G_0J) - \bar{\mathcal V}[G_0J],\label{Seq:Gc_V}
\end{align}
where we define \(\bar{\mathcal V}[\psi]\) as 
\begin{align}
    \bar{\mathcal V}[\psi] \equiv - \ln\langle e^{-V[\phi+\psi]}\rangle_{S_0[\phi]}.\label{Seq:eff_int}
\end{align}
From Eq.~\eqref{Seq:Gc_V}, it is now clear that the connected Green functions have the power counting
\begin{align}
    G_{{\rm c},\alpha_1\alpha_2}^{(2)} &= [G_0]_{\alpha_1\alpha_2} + \mathcal O(G_0^2),\label{Seq:Gc2_pc}\\
    G_{{\rm c},\alpha_1\cdots\alpha_n}^{(n)} &= \mathcal O(G_0^{n})\ \ (n\neq2).\label{Seq:Gcn_pc}
\end{align}
In particular, in the zero-fluctuation limit \(G_0\to 0\), Eqs.~\eqref{Seq:Gc2_pc} and \eqref{Seq:Gcn_pc} prove that the leading contribution in the correlation functions \(G_{\alpha_1\cdots\alpha_n}^{(n)}\) originates from the product of two-point correlations \(G_{{\rm c},\alpha_i\alpha_j}^{(2)}\), such as \(G_{{\rm c},\alpha_1\alpha_2}^{(2)} G_{{\rm c},\alpha_3\alpha_4}^{(2)}\cdots G_{{\rm c},\alpha_{n-1}\alpha_n}^{(2)}\). In other words, the correlation functions asymptotically converge to the ones obtained in the quadratic model \(S_0[\phi]=\frac{1}{2}(\phi,G_0^{-1}\phi)\). We note that this discussion rigorously proves the scale separation property of the RG in Eq.~(3) in the main text.

From the definition in Eq.~\eqref{Seq:eff_int}, \(\bar{\mathcal V}[\psi]\) can be interpreted as an effective interaction, which incorporates the fluctuations mediated by the Gaussian propagator \(G_0\) to the original interaction \(V[\phi]\). In fact, this interpretation becomes clearer by formally rewriting \(\exp(-\bar{\mathcal V}[\psi])\) as 
\begin{align}
    e^{-\bar{\mathcal V}[\psi]} &= \frac{1}{Z_0} \int [d\phi]\ e^{-S_0[\phi]} e^{-V[\phi+\psi]}\\
    &= \frac{1}{Z_0} \int [d\phi]\ e^{-S_0[\phi]} e^{\phi\frac{\delta}{\delta\psi}} e^{-V[\psi]}\\
    &= e^{\frac{1}{2}\left(\frac{\delta}{\delta\psi},G_0\frac{\delta}{\delta\psi}\right)} e^{-V[\psi]}.\label{Seq:V_formal}
\end{align}
Here, we use the translation formula \(e^{\phi\frac{\delta}{\delta \psi}}f[\psi] = f[\psi+\phi]\) and perform the Gaussian integral as \(\int[d\phi]\, e^{-S_0[\phi]}e^{\phi\frac{\delta}{\delta\psi}}=Z_0e^{\frac{1}{2}\left(\frac{\delta}{\delta\psi},G_0\frac{\delta}{\delta\psi}\right)}\). Equation~\eqref{Seq:V_formal} clearly shows how the Gaussian propagator \(G_0\) changes the original interaction \(V[\phi]\) to \(\bar{\mathcal V}[\phi]\). 

\subsubsection{\label{Sss:Polc}Derivation of the Polchinski RG equation}
\begin{figure}
    \centering
    \includegraphics[width=5cm]{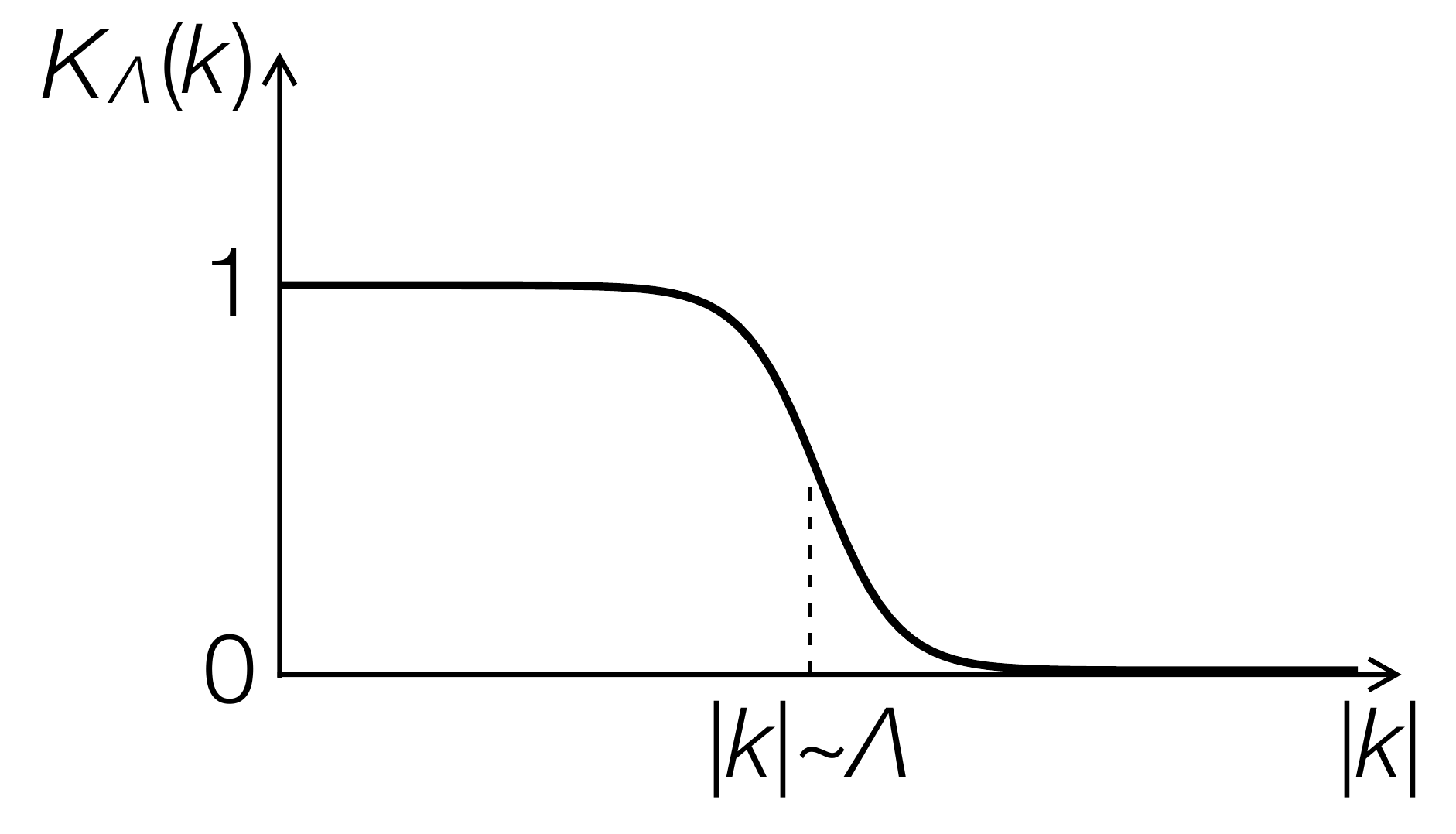}
    \caption{Schematic of \(K_\Lambda(k)\), whose value smoothly changes from one to zero around \(|k|\!\sim\!\Lambda\).\label{Sfig:KL}}
\end{figure}
In the original paper \cite{polchinski1984a}, Polchinski revisited the concept of renormalizability in high-energy physics by deriving the exact RG flow equation. 
Based on the functional representation of the correlation functions described above, here we construct the exact effective model of in Eq.~\(S[\phi]\)~\eqref{Seq:S} and summarize the derivation of the Polchinski RG equation.  
To this end, we first decompose the Gaussian propagator \(G_0\) into two parts: the Gaussian propagators above and below the RG scale \(\Lambda\) as 
\begin{align}
    G_0 &= G_{0\Lambda}^< + G_{0\Lambda}^>,\\
    G_{0\Lambda}^{<}(k) &= K_\Lambda(k) G_{0}(k),\\
    G_{0\Lambda}^{>}(k) &= (1-K_\Lambda(k)) G_{0}(k),
\end{align}
where \(K_\Lambda(k)\) is the RG cutoff function (cf. Fig.~\ref{Sfig:KL}). 
As demonstrated in the previous section, a model in the form \(S_\Lambda[\phi]=\frac{1}{2}(\phi,G_{0\Lambda}^{<-1}\phi) + V_{\Lambda}[\phi]\) satisfies the scale-separation property of the RG (Eq.~(3) in the main text). 
We construct the interaction \(V_\Lambda[\phi]\) that preserves the correlations of the UV model \(S[\phi]\) up to the RG scale \(\Lambda\) by rewriting the effective interaction \(\bar{\mathcal V}[\psi]\) in Eq.~\eqref{Seq:eff_int} of the original model \(S[\phi]\) as (cf. Eq.~\eqref{Seq:V_formal})
\begin{align}
    e^{-\bar{\mathcal V}[\psi]} &= 
    e^{\frac{1}{2}\left(\frac{\delta}{\delta\psi},G_{0\Lambda}^<\frac{\delta}{\delta\psi}\right)} 
    e^{\frac{1}{2}\left(\frac{\delta}{\delta\psi},G_{0\Lambda}^>\frac{\delta}{\delta\psi}\right)} 
    e^{-V[\psi]}\\
    &\equiv
    e^{\frac{1}{2}\left(\frac{\delta}{\delta\psi},G_{0\Lambda}^<\frac{\delta}{\delta\psi}\right)}
    e^{-{\bar{\mathcal V}}_\Lambda[\psi]}.\label{Seq:V_formal_sep}
\end{align}
Here, we define \(\bar{\mathcal V}_\Lambda[\psi]\) by \(e^{-{\bar{\mathcal V}}_\Lambda[\psi]}=e^{\frac{1}{2}\left(\frac{\delta}{\delta\psi},G_{0\Lambda}^>\frac{\delta}{\delta\psi}\right)}e^{-V[\psi]}\), which incorporates the fluctuations above the RG scale \(\Lambda\) mediated by \(G_{0\Lambda}^>\). 
From Eqs.~\eqref{Seq:V_formal} and \eqref{Seq:V_formal_sep}, it is now clear that the effective interaction of the original model \(\bar{\mathcal V}[\psi]\) can also be regarded as the effective interaction of the model  
\begin{align}
    S_\Lambda[\phi] &= \frac{1}{2}(\phi, G_{0\Lambda}^<\phi) + \bar{\mathcal V}_\Lambda[\phi].\label{Seq:SL_cand}
\end{align}
In particular, from Eq.~\eqref{Seq:Gc_V}, the generating functional for the connected Green functions of the model \(S_\Lambda[\phi]\) is given by 
\begin{align}
    \mathcal G_{c\Lambda}^<[J] &= \frac{1}{2}(J, G_{0\Lambda}^<J) - \bar{\mathcal V}[G_{0\Lambda}^<\Lambda J]\\
    &= \frac{1}{2}(J, K_\Lambda G_0J) - \bar{\mathcal V}[K_\Lambda G_0J].\label{Seq:GcL_gen}
\end{align}
Since \(K_\Lambda(k)= 1\) at \(|k|<\Lambda\), Eq.~\eqref{Seq:GcL_gen} proves that the connected Green functions in \(S_\Lambda[\phi]\) below the RG scale \(\Lambda\) are identical to the ones in the original model \(S[\phi]\). Since the correlation functions can be obtained through the connected Green functions as in Eqs.~\eqref{Seq:corr_decomp1}-\eqref{Seq:corr_decomp3}, \(S_\Lambda[\phi]\) exactly preserves all the correlations of the original model below the RG scale \(\Lambda\). In other words, \(S_\Lambda[\phi]\) is the exact effective model of \(S[\phi]\) at RG scale \(\Lambda\). 

\begin{itembox}{Exact effective model}
    Effective model of \(S[\phi]= \frac{1}{2}(\phi,G_0^{-1}\phi)+V[\phi]\) that exactly preserves all the correlations below the RG scale \(\Lambda\) is given by
    \begin{align}
        S_\Lambda[\phi] &= \frac{1}{2}(\phi,G_{0\Lambda}^{-1}\phi) + V_\Lambda[\phi],\label{Seq:s:SL}\\
        G_{0\Lambda}(k) &= K_\Lambda(k) G_0(k),\label{Seq:s:G0L}\\
        e^{-V_\Lambda[\phi]} &= e^{\frac{1}{2}\left(\frac{\delta}{\delta\phi}(G_0-G_{0\Lambda})\frac{\delta}{\delta\phi}\right)} e^{-V[\phi]}.\label{Seq:s:VL}
    \end{align}
\end{itembox}

We are now in a position to prove the Polchinski RG flow equation (Eq.~(6) in the main text), which is the flow equation of \(V_\Lambda[\phi]\) in Eq.~\eqref{Seq:s:VL}. 
We note that \(V_\Lambda[\phi]\) at the UV scale \(\Lambda=\Lambda_{\rm UV}\) coincides with the original interaction \(V[\phi]\), which is clear by noting that the differential operator in Eq.~\eqref{Seq:s:VL} at \(\Lambda=\Lambda_{\rm UV}\) does nothing to its right side \(e^{-V[\phi]}\), since \(V[\phi]\) has the wavenumber cutoff at \(\Lambda=\Lambda_{\rm UV}\).
The Polchinski equation is then obtained by differentiating both sides of Eq.~\eqref{Seq:s:VL} as 
\begin{align}
    - e^{-V_{\Lambda}[\phi]} \partial_\Lambda V_{\Lambda}[\phi] &= -\frac{1}{2}\left(\frac{\delta}{\delta\phi}, \partial_\Lambda G_{0\Lambda}\frac{\delta}{\delta\phi}\right) e^{-V_{\Lambda}[\phi]}\\
    &= e^{-V_{\Lambda}[\phi]}\left[
        \frac{1}{2}\left(\frac{\delta}{\delta\phi}, G_0\partial_\Lambda K_{\Lambda}\frac{\delta}{\delta\phi}\right) V_{\Lambda}[\phi]
        - \frac{1}{2}\left(\frac{\delta V_{\Lambda}[\phi]}{\delta\phi}, \partial_\Lambda G_{0\Lambda}\frac{\delta V_{\Lambda}[\phi]}{\delta\phi}\right)
        \right],
\end{align}
which proves the following:

\begin{itembox}{Polchinski RG flow equation}
    The flow equation of \(V_\Lambda[\phi]\) in Eq.~\eqref{Seq:s:VL} is 
    \begin{align}
        \partial_\Lambda V_{\Lambda}[\phi] = 
        -\frac{1}{2}\left(\frac{\delta}{\delta\phi}, \partial_\Lambda G_{0\Lambda}\frac{\delta}{\delta\phi}\right) V_{\Lambda}[\phi]
        + \frac{1}{2}\left(\frac{\delta V_{\Lambda}[\phi]}{\delta\phi}, \partial_\Lambda G_{0\Lambda}\frac{\delta V_{\Lambda}[\phi]}{\delta\phi}\right),\label{Seq:s:Pol}
    \end{align}
    where the initial condition of the flow is \(V_{\Lambda_{\rm UV}}[\phi]\!=\!V[\phi]\).
\end{itembox}
We note that Eq.~\eqref{Seq:s:Pol} can be explicitly written as
\begin{align}
    \partial_\Lambda V_{\Lambda}[\phi] = -\frac{1}{2}\int\frac{d^dk}{(2\pi)^d}\ G_{0}(k)\partial_\Lambda K_{\Lambda}(k) \left(\frac{\delta^2 V_{\Lambda}[\phi]}{\delta\phi_k\delta\phi_{-k}}
    - \frac{\delta V_{\Lambda}[\phi]}{\delta\phi_k}\frac{\delta V_{\Lambda}[\phi]}{\delta\phi_{-k}} \right),\label{Seq:der_Pol}
\end{align}
which reproduces Eq.~(6) in the main text with \(G_0(k)=k^{-2}\). 
Here, we define the Fourier transform of the functional derivative as 
\begin{align}
    \frac{\delta}{\delta\phi(x)} &\equiv \int \frac{d^dk}{(2\pi)^d} e^{-ikx} \frac{\delta}{\delta\phi_k},\ \ 
    \frac{\delta}{\delta\phi_k} \equiv \int d^dx\ e^{ikx} \frac{\delta}{\delta\phi(x)},
\end{align}
which reads \(\delta\phi(x)/{\delta\phi(x')}=\delta(x-x')\) and \(\delta\phi_k/\delta\phi_{k'}=(2\pi)^d\delta(k-k')\).

\subsubsection{\label{Sss:diff}Derivation of the convex-diffusion equation for the probability distribution during the RG flow}
We here provide the derivation of the convex-diffusion equation in Eq. (7), which is the RG flow equation for the probability functional. 
To this end, we first derive the flow equation of the distribution \(p_\Lambda[\phi] \propto e^{-S_\Lambda[\phi]}\) based on the Polchinski RG flow equation~\eqref{Seq:s:Pol}. After that, we derive the flow equation for the rescaled distribution \(p'_\Lambda[\phi] \propto e^{-S_\Lambda[\sqrt{K_\Lambda}\phi]}\), which gives the basis of the renormalization group diffusion model (RGDM)  developed in the main text. 

 We start from rewriting \(\partial_\Lambda e^{-S_\Lambda[\phi]}\) as \cite{cotler2023b} 
\begin{align}
    \partial_\Lambda e^{-S_\Lambda[\phi]} 
    &= \frac{1}{2}{\rm Tr}\left[\frac{\partial_\Lambda G_{0\Lambda}}{G_{0\Lambda}}\right]e^{-S_\Lambda[\phi]}
    - \frac{1}{2}\left(\frac{\delta}{\delta\phi} \partial_\Lambda G_{0\Lambda} \frac{\delta}{\delta\phi}\right) e^{-S_\Lambda[\phi]}
    - \left(\frac{\delta}{\delta\phi}, \frac{\partial_\Lambda G_{0\Lambda}}{G_{0\Lambda}}\,\phi\, e^{-S_\Lambda[\phi]}\right).\label{Seq:der_eSL}
\end{align}
To derive this, we use the Polchinski RG flow equation~\eqref{Seq:s:Pol} to rewrite \(e^{-S_\Lambda[\phi]}\partial_\Lambda V_{\Lambda}[\phi]\) as 
\begin{align}
    e^{-S_\Lambda[\phi]}\partial_\Lambda V_{\Lambda}[\phi]
    &= e^{-S_\Lambda}\left[-\frac{1}{2}\left(\delta, \partial_\Lambda G_{0\Lambda}\delta\right) (S_{\Lambda} - S_{0\Lambda})
    + \frac{1}{2}\left(\delta (S_{\Lambda}-S_{0\Lambda}), \partial_\Lambda G_{0\Lambda} \delta(S_{\Lambda}-S_{0\Lambda})\right)\right]\\
    &= \frac{1}{2}(\delta,\partial_\Lambda G_{0\Lambda} \delta) e^{-S_\Lambda} 
    + \frac{1}{2} e^{-S_\Lambda}\left[(\delta, \partial_\Lambda G_{0\Lambda} \delta) S_{0\Lambda} 
    + (\delta S_{0\Lambda}, \partial_\Lambda G_{0\Lambda} \delta S_{0\Lambda})
    - 2 (\delta S_{0\Lambda}, \partial_\Lambda G_{0\Lambda} \delta S_{\Lambda})\right].\label{Seq:Pol_der1}
\end{align}
Here, \(S_0[\phi]=\frac{1}{2}(\phi, G_{0\Lambda}^{-1}\phi)\), and we abbreviate the model \(S_{(0)}[\phi]\) and the derivative \(\frac{\delta}{\delta\phi}\) by \(S_{(0)}\) and \(\delta\), respectively. From \(\delta S_0 = G_{0\Lambda}^{-1}\phi\), we further simplify Eq.~\eqref{Seq:Pol_der1} to
\begin{align}
    e^{-S_\Lambda[\phi]}\partial_\Lambda V_{\Lambda}[\phi] &= 
    \frac{1}{2}(\delta,\partial_\Lambda G_{0\Lambda} \delta) e^{-S_\Lambda} 
    + \frac{1}{2} e^{-S_\Lambda}\left[
        {\rm Tr}\left[\frac{\partial_\Lambda G_{0\Lambda}}{G_{0\Lambda}}\right]
        - 2\partial_\Lambda S_{0\Lambda}
        - 2\left(\phi, \frac{\partial_\Lambda G_{0\Lambda}}{G_{0\Lambda}} \delta S_\Lambda\right)
    \right]\\
    &= \frac{1}{2}(\delta,\partial_\Lambda G_{0\Lambda} \delta) e^{-S_\Lambda} 
    - \frac{1}{2} e^{-S_\Lambda}{\rm Tr}\left[\frac{\partial_\Lambda G_{0\Lambda}}{G_{0\Lambda}}\right]
    - e^{-S_\Lambda} \partial_\Lambda S_{0\Lambda}
    + \left(\delta, \frac{\partial_\Lambda G_{0\Lambda}}{G_{0\Lambda}}\,\phi\, e^{-S_\Lambda}\right).\label{Seq:Pol_der2}
\end{align}
Since \(\partial_\Lambda e^{-S_\Lambda[\phi]}=e^{-S_\Lambda[\phi]}\partial_\Lambda(S_{0\Lambda}[\phi]+V_{\Lambda}[\phi])\), Eq.~\eqref{Seq:Pol_der2} proves Eq.~\eqref{Seq:der_eSL}.
Also, by integrating Eq.~\eqref{Seq:der_eSL} with respect to the field \(\phi\), we obtain the following formula for the distribution function \(Z_\Lambda = \int[d\phi] e^{-S_{\Lambda}[\phi]}\) as
\begin{align}
    \partial_\Lambda Z_\Lambda &= \frac{1}{2} {\rm Tr}\left[\frac{\partial_\Lambda G_{0\Lambda}}{G_{0\Lambda}}\right] Z_\Lambda.\label{Seq:der_ZL}
\end{align}
Here, we drop the path-integral of total derivative terms. 
With Eqs.~\eqref{Seq:der_eSL} and \eqref{Seq:der_ZL}, we arrive at the following expression for the derivative of \(p_\Lambda[\phi]=e^{-S_\Lambda[\phi]}/Z_\Lambda\) as \cite{cotler2023b}
\begin{align}
    \partial_\Lambda p_\Lambda[\phi] &= \partial_\Lambda 
    \left(\frac{e^{-S_\Lambda[\phi]}}{Z_\Lambda}\right)\\
    &= - \frac{1}{2}\left(\frac{\delta}{\delta\phi} \partial_\Lambda G_{0\Lambda} \frac{\delta}{\delta\phi}\right) p_\Lambda[\phi]
    - \left(\frac{\delta}{\delta\phi}, \frac{\partial_\Lambda G_{0\Lambda}}{G_{0\Lambda}} \phi\ p_\Lambda[\phi]\right).\label{Seq:der_PL_1}
\end{align}

\begin{figure}
    \centering
    \includegraphics[width=14cm]{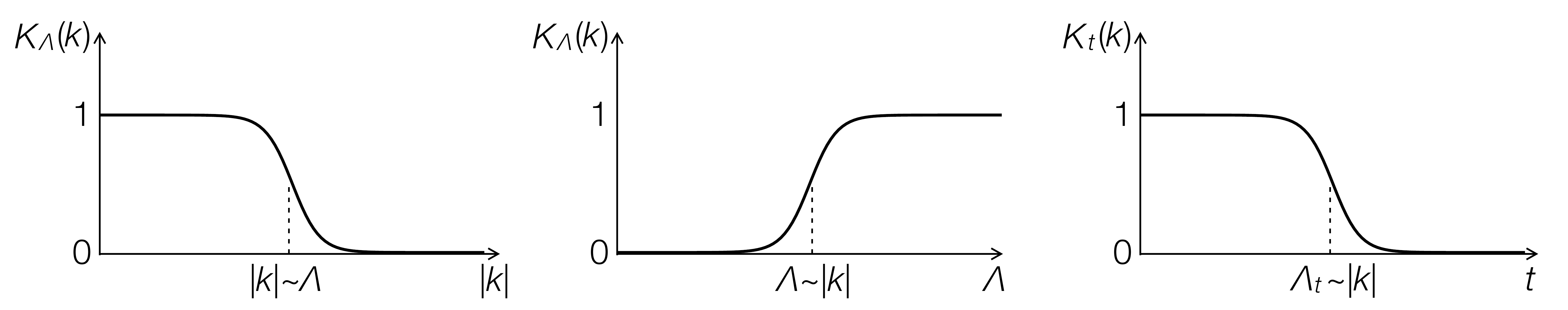}
    \caption{\label{Sfig:Kt} Schematic of \(K_\Lambda(k)\) as a function of \(k, \Lambda,\) and \(t\), where \(t\propto -\ln\Lambda\) is the logarithmic RG time.}
\end{figure}

We next derive the flow equation for the rescaled distribution \(p'_\Lambda[\phi]\propto e^{-S[\sqrt{K_\Lambda}\phi]}\). To this end, we first express \(p'_\Lambda[\phi]\) as  
\begin{align}
    p'_\Lambda[\phi] = c_\Lambda\, p_\Lambda[K_\Lambda^{\frac{1}{2}}\phi],
\end{align}
where the constant \(c_\Lambda\) is given by \(c_\Lambda\!=\!\sqrt{\prod_k K_\Lambda(k)}\) to ensure the normalization condition \(\int[d\phi]\,p'_\Lambda[\phi]\!=\!1\). We then rewrite \(\partial_\Lambda p'_\Lambda[\phi]\) as
\begin{align}
    \partial_\Lambda p'_\Lambda[\phi] 
    & = (\partial_\Lambda c_\Lambda) p_\Lambda[K_\Lambda^{\frac{1}{2}}\phi]
    + c_\Lambda (\partial_\Lambda p_\Lambda[\phi])\Bigr|_{\phi=K_\Lambda^{\frac{1}{2}}\phi}
    + c_\Lambda \left(\partial_\Lambda K_\Lambda^{\frac{1}{2}}\,\phi, \frac{\delta p_\Lambda[\phi]}{\delta \phi}\right)\biggr|_{\phi=K_\Lambda^{\frac{1}{2}}\phi}.\label{Seq:der_PL_2}
\end{align}
With Eqs.~\eqref{Seq:der_PL_1} and \eqref{Seq:der_PL_2} and the formulae
\begin{align}
    \frac{\delta p_\Lambda[\phi]}{\delta\phi}\biggr|_{\phi=K_\Lambda^{\frac{1}{2}}\phi} &= c_\Lambda^{-1} K_\Lambda^{-\frac{1}{2}} \frac{\delta p'_\Lambda[\phi]}{\delta\phi},\\
    \left[\left(\frac{\delta}{\delta\phi},\partial_\Lambda G_{0\Lambda}\frac{\delta}{\delta\phi}\right)p_{\Lambda}[\phi] \right]\biggr|_{\phi=K_\Lambda^{\frac{1}{2}}\phi} 
    &= c_\Lambda^{-1} \left(\frac{\delta}{\delta\phi},G_0 \frac{\partial_\Lambda K_\Lambda}{K_\Lambda}\frac{\delta}{\delta\phi}\right)p'_{\Lambda}[\phi],
\end{align}
we finally arrive at 
\begin{align}
    \partial_\Lambda p'_\Lambda[\phi] &= -\frac{1}{2}\left(\frac{\delta}{\delta\phi},G_0 \frac{\partial_\Lambda K_\Lambda}{K_\Lambda}\frac{\delta}{\delta\phi}\right)p'_{\Lambda}[\phi]
    - \frac{1}{2}\left(\frac{\delta}{\delta\phi}, \frac{\partial_\Lambda K_\Lambda}{K_\Lambda}\,\phi\,p'_\Lambda[\phi]\right).\label{polscaled}
\end{align}
Here, we also use the following expression of \(\partial_\Lambda c_\Lambda\) as 
\begin{align}
    \partial_\Lambda c_\Lambda = \partial_\Lambda \exp\left(\frac{1}{2}{\rm Tr}\ln K_\Lambda\right) = \frac{c_\Lambda}{2}{\rm Tr}\left[\frac{\partial_\Lambda K_\Lambda}{K_\Lambda}\right].
\end{align}
In terms of the logarithmic RG `time' scale \(t\), which is defined by the relation \(\Lambda=\Lambda_0 e^{-t/\tau}\), the evolution of the probability functional in Eq.~\eqref{polscaled} becomes 
\begin{align}
    \partial_t p_t[\phi] &= -\frac{1}{2}\left(\frac{\delta}{\delta\phi} G_0 \frac{\partial_t K_t}{K_t} \frac{\delta}{\delta\phi}\right) p_t[\phi]
    - \frac{1}{2}\left(\frac{\delta}{\delta\phi}, \frac{\partial_t K_t}{K_t}\, \phi\, p_t[\phi]\right)\\
    &= -\frac{1}{2}\int \frac{d^dk}{(2\pi)^d}\, G_0(k)\frac{\partial_t K_t(k)}{K_t(k)} \frac{\delta^2 p_t[\phi]}{\delta\phi_k\delta\phi_{-k}} 
    - \int \frac{d^dk}{(2\pi)^d} \frac{\delta}{\delta\phi_{k}}\left(\frac{1}{2}\frac{\partial_t K_t(k)}{K_t(k)}\phi_{k} p_t[\phi]\right),\label{Seq:FP}
\end{align}
which is Eq.~(7) in the main text with \(G_0(k)=k^{-2}\). Here, we abbreviate \(p_{\Lambda_t}[\phi]\) and \(K_{\Lambda_t}\) by \(p_t[\phi]\) and \(K_t\), respectively. 
We recall that \(K_t(k)\) is a monotonically decreasing function of \(t\) and satisfies \(\partial_t K_t(k)<0\) (see, e.g., Fig.~\ref{Sfig:Kt}). Therefore, Eq.~\eqref{Seq:FP} can be regarded as a convex-diffusion equation with wavenumber-dependent diffusion coefficient \(g_{tk}\!=\!-G_0(k)\partial_tK_t(k)/K_t(k)\!>\!0\) and drift term \(f_{tk}\!=\!(\partial_tK_t(k)/2K_t(k))\phi_k\). In particular, the corresponding Langevin equation for the field \(\phi\) is given by 
\begin{align}
    d\phi_{tk} &= \frac{1}{2}\frac{\partial_t K_t(k)}{K_t(k)}\phi_{tk} dt + \sqrt{-G_0(k)\frac{\partial_tK_t(k)}{K_t(k)}}dw_k,
\end{align}
where \(dw_k\) is the Wienner process defined by \(\mathbb E[dw_{k'}dw_k]=\delta_{kk'}dt\). As is known in the theory of stochastic differential equation, this Langevin equation has the conditional probability
\begin{align}
    p(\phi_t|\phi_0) &= \prod_k \mathcal N(\phi_{tk}|\sqrt{\bar\alpha_{tk}}\phi_{0k},\bar{\beta}_{tk}),\\
    \sqrt{\bar\alpha_{tk}} &= \exp\left(\frac{1}{2}\int_{0}^{t} ds\,\frac{\partial_s K_s(k)}{K_s(k)}\right) = \sqrt{K_t(k)}, \\
    \bar\beta_{tk} &= \bar\alpha_{tk} \int_0^t ds\, \bar\alpha_{sk}^{-1}\left(-G_0(k)\frac{\partial_sK_s(k)}{K_s(k)}\right) = G_{0}(k)(1-K_t(k)),
\end{align}
which proves Eqs.~(8) and (9) in the main text. We here use \(K_{t=0}(k)\!=\!1\), which follows from our choice of the RG scale \(\Lambda_t\) as \(\Lambda_{0}\!=\!\Lambda_{\rm UV}\) \cite{footnote1}. 

\subsection{\label{Ss:Tech_RGDM}Technical details of the renormalization group diffusion model}
We provide the technical details of the renormalization group diffusion model (RGDM). Since data of the protein structure and image are defined on a finite-size lattice instead of the Euclidian space as considered in the previous chapter, we first provide a way to apply the RG framework above to the discrete space and summarize the notations used for this in Sec.~\ref{Sss:sum_not}. We then provide the details of the training and sampling schemes in the RGDM in Sec.~\ref{Sss:tra_samp}. The values of the hyperparameters and the structure of the deep neural networks used in the numerical experiments are given in \ref{Sss:nn_param} and Secs.~\ref{Sss:ns_param}. Also, we provide typical samples generated by the diffusion models in our numerical experiments in Sec.~\ref{Sss:sampled_data}.

\subsubsection{\label{Sss:sum_not}Summary of notations}
\begin{figure}
    \centering
    \includegraphics[width=10cm]{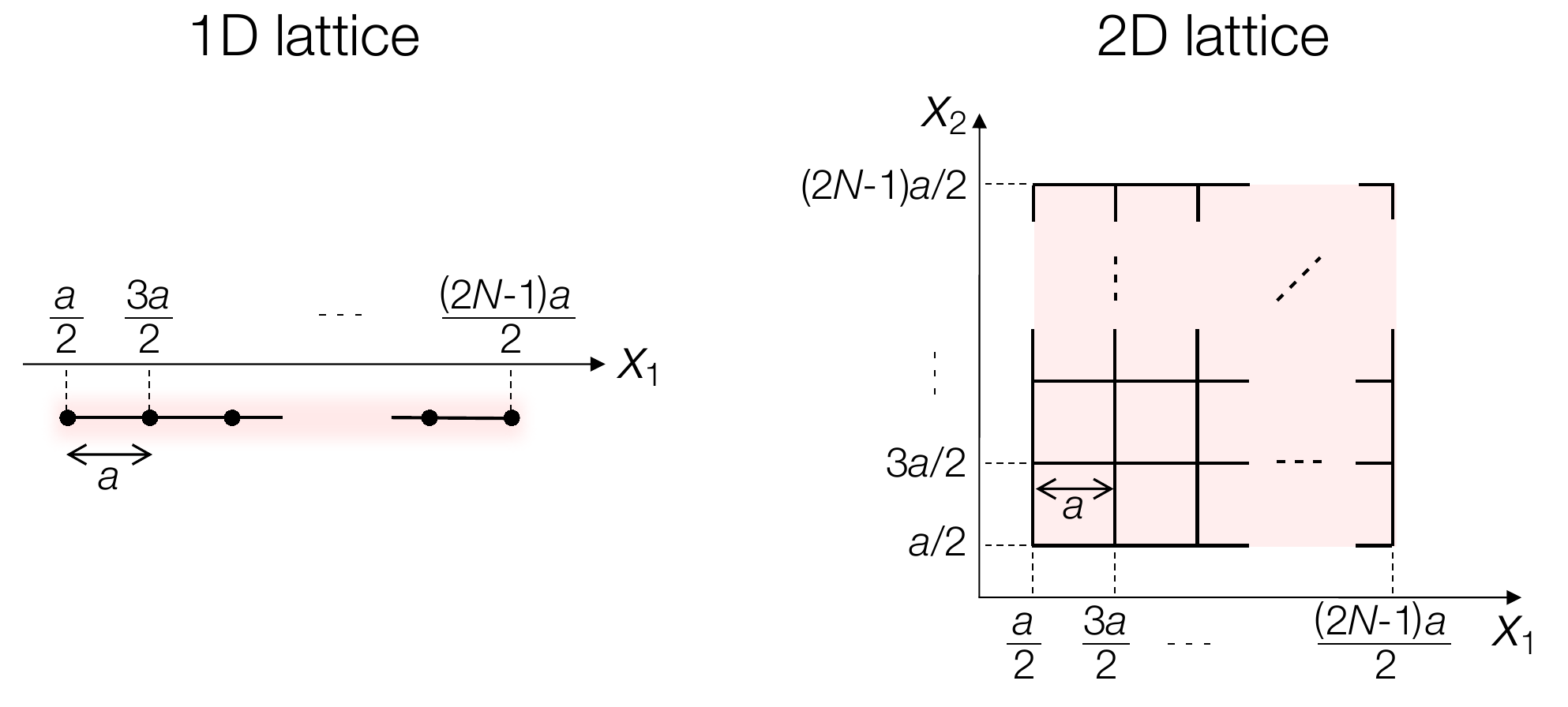}
    \caption{\label{Sfig:lattice} The coordinate of \(d\)-dimensional lattice with \(N^d\) sites for \(d=1,2\).}
\end{figure}

As described in the main text, we consider data that can be regarded as a field on a \(d\)-dimensional space. To be concrete, we consider a finite-size hypercubic lattice with \(N^d\) sites, in which the lattice coordinate \(x\) takes \(x=\left(\left(n_1+1/2\right)a,\ldots,\left(n_d+1/2\right)a\right)^T\) \((n_i = 0, \ldots, N-1)\) with lattice spacing \(a\) (Fig.~\ref{Sfig:lattice}). We define the Fourier modes \(\phi_k\) by the orthonormal discrete cosine transform (DCT) of \(\phi(x)\) as 
\begin{align}
    \phi_k = \left\{
        \begin{array}{ll}
            \sqrt{\frac{2}{N^d}} \sum_{x} \phi(x) \cos(kx) & (k\neq 0)\\
            \sqrt{\frac{1}{N^d}} \sum_{x} \phi(x) & (k=0),
        \end{array}
    \right.\label{Seq:DCT}
\end{align}
where the wavenumber \(k\) takes a discrete vector value \(k=\left(\frac{n_1\pi}{Na},\ldots,\frac{n_d\pi}{Na}\right)^{\rm T}\) with \(n_i\in\{0, \ldots, N-1\}\). Here, we use the DCT instead of the Fourier transform for just a practical reason; since \(\phi_k\) and the transform matrix from \(\phi(x)\) to \(\phi_k\) are real-valued in the DCT, it is more amenable to implementations by the PyTorch libraries in Python. Below we take the unit of \(a=1\).

The model ansatz of natural data (Eq.~(2) in the main text) becomes 
\begin{align}
    S_{\rm data}(\phi) &= \frac{1}{2}\sum_k v^{-1}k^2\phi_k^2 + V(\phi),
\end{align}
where \(v\) is a constant that depends on how one chooses the unit of the field \(\phi(x)\).
Here, we note that \(S_{\rm data}(\phi)\) naturally has both IR and UV cutoffs at \(\Lambda_{\rm IR} = \pi/N\) and \(\Lambda_{\rm UV} = (N-1)\pi/N \simeq \pi\), respectively. 
Meanwhile, the RG flow used in the RGDM (Eq.~\eqref{Seq:FP} and Eq.~(7) in the main text) becomes 
\begin{align}
    \partial_t p_t(\phi) = 
    - \frac{1}{2} \sum_k \frac{v}{k^2}\frac{\partial_t K_t(k)}{K_t(k)} \frac{\partial^2 p_t(\phi)}{\partial^2\phi_k}
    - \sum_k \frac{\partial}{\partial\phi_k} \left(\frac{1}{2}\frac{\partial_t K_t}{K_t}\phi_k p_t(\phi)\right).\label{Seq:rgdm-FP}
\end{align}
The forward and backward diffusions at \(t=1,\cdots,T\) are then given as (cf. Eqs.~(8) in the main text and Eq.~(11) in the Method)
\begin{align}
    \text{forward}: \phi_{tk} &= \sqrt{\bar\alpha_{tk}}\phi_{0k} + \sqrt{\bar\beta_{tk}}\epsilon_k,\label{Seq:fdiff}\\
    \text{backward}: \phi_{t-1k} &= \frac{1}{\sqrt{\alpha_{tk}}}\left(\phi_{tk} - \frac{\beta_{tk}}{\bar\beta_{tk}}\xi_{\theta k}(\phi_t,t)\right) + \sqrt{\beta_{tk}}\epsilon_k,\label{Seq:bdiff}
\end{align}
respectively. 
In the numerical experiments, we introduce a small parameter \(m\) and replace \(k^2\) in the noise schedule by \(k^2+m^2\) to avoid divergent singularity around \(k\sim 0\). In our work, we take \(v\) and \(m\) so that the variance of the dataset \({\rm Var}[\phi_k]\) roughly matches \(v/(k^2+m^2)\).
Altogether, the noise schedule of the RGDM is given by
\begin{align}
    \bar\alpha_{tk} &= K_t(k),\ \ 
    \alpha_{tk} = \left\{
        \begin{array}{ll}
            K_t(k) & (t=1)\\
            K_t(k)/K_{t-1}(k) & (t=2,\ldots,T)
        \end{array}
        \right.,
        \label{Seq:param:a}\\
    \bar\beta_{tk} &= \frac{v}{k^2+m^2} (1-\bar\alpha_{tk}),\ \ 
    \beta_{tk} = \frac{v}{k^2+m^2} (1-\alpha_{tk}),\label{Seq:param:b}\\
    K_t(k) &= \frac{r_{tk}}{1+r_{tk}},\ \ r_{tk} = r\left(k^2/\Lambda_t^2\right),\ \ \Lambda_t = \Lambda_0 e^{-t/\tau}.\label{Seq:param:Lam}
\end{align}
Under this noise schedule, any data distribution \(p_{\rm data}\) at a UV scale converges to the Gaussian distribution \(p_{\rm GS}(\phi)=\prod_k\mathcal N(\phi_k;0,v/(k^2+m^2))\) in the IR limit \(t\to\infty\).

In the RGDM, we introduce the RG projection layers to discard higher wavenumber components, which are integrated out in the RG. Specifically, the RG projection layers at the \(t\)-th diffusion step receives \(\phi_t(x)\), i.e., the field represented in real space, and returns the real-space field \((\mathcal P_t\phi_t)(x)\) with DCT components
\begin{align}
    (\mathcal P_t\phi_t)_k &= \left\{
        \begin{array}{cc}
            \phi_{tk} & (|k|\leq c\Lambda_t)\\
            0 & (|k| > c\Lambda_t)
        \end{array}
    \right..\label{Seq:RG_proj}
\end{align}
To determine the constant \(c\), we first introduce a small wavenumber parameter \(k_{\rm cutoff}\sim \Lambda_{\rm IR}\) and determine the value of $c$ so that only \(\phi_{k}\) with \(|k|\leq k_{\rm cutoff}\) are retained at \(t=T\). Thus, we take \(c\) to be
\begin{align}
    c\Lambda_T = k_{\rm cutoff}.\label{Seq:k_cutoff}
\end{align}
As shown in Fig.~1{e} in the main text, the RG projection layers are added before and after the denoising deep neural network (DNN), such as UNet. During the training of the DNN, we use the cost function as (cf. Eq.(10) in the Method)
\begin{align}
    L_\theta = \sum_{t=1}^T \lambda_t\ \mathbb E_{\phi_0\sim p_0(\phi),\xi_t\sim\mathcal{N}(0,\bar\beta_t)}
    \left[||\xi_\theta(\phi_t,t)-\xi_t||^2\right].\label{Seq:cost}
\end{align}
In Eq.~\eqref{Seq:cost}, we take the time-dependent weight \(\lambda_t\) as 
\begin{align}
    \lambda_t &= \frac{D}{\sum_{k\leq c\Lambda_t}\bar\beta_{tk}}
\end{align}
so that each term in \(L_\theta\) would have approximately the same weight. In our work, we take the overall factor \(D\) as \(D=\sum_{k}\bar\beta_{t=\infty k}=\sum_k v/(k^2+m^2)\).

\subsubsection{\label{Sss:tra_samp}Training and sampling schemes}
As described in the main text, we simplify the training and sampling schemes in the RGDM by using the RG projection layers, which we detail here.
For the sake of convenience, here we use the notation where the plain Greek letters such as \(\phi\) and \(\epsilon\) represent the field in real space, while the Greek letters with a bar such as \(\tilde\phi\) and \(\tilde\epsilon\) represent the field in wavenumber space. Also, we denote the DCT transform and its inverse by \({\rm DCT}(\phi)\) and \({\rm DCT}^{-1}(\tilde\phi)\), respectively. 

Firstly, the RG projection is defined by the following algorithm (cf. Eq.~\eqref{Seq:RG_proj}) \cite{footnote1}:
\begin{table}[h]
    \centering
    \begin{tabular}{ll}\hline
      \multicolumn{2}{l}{\textbf{Algorithm 1}: RG projection function\hspace*{130mm}}\\\hline
      \multicolumn{2}{l}{\, input: \((\phi,t)\)}\\
      \, 1: & \(\tilde\phi = {\rm DCT}(\phi)\)\\
      \, 2: & \textbf{for} \(|k|>c\Lambda_t\) \textbf{do}\\
      \, 3: & \quad \(\tilde\phi_k:=0\)\\
      \, 4: & \textbf{end for}\\
      \, 5: & \(\phi^< = {\rm DCT}^{-1}(\tilde\phi)\)\\
      \, 6: & \textbf{return} \(\phi^<\)\\\hline 
    \end{tabular}
\end{table}

\noindent
With the RG projection function defined above, the training of the RGDM can be summarized as follows (cf. Eqs.\eqref{Seq:fdiff}, \eqref{Seq:param:a}, \eqref{Seq:param:b}, and \eqref{Seq:cost}, and Fig.~1e in the main text.):
\begin{table}[h]
    \centering
    \begin{tabular}{rl}\hline
      \multicolumn{2}{l}{\textbf{Algorithm 2}: Training scheme of the RGDM\hspace*{120mm}}\\\hline
      \, 1: & \textbf{repeat}\\
      \, 2: & \quad \(\phi_0\sim p_{\rm data}(\phi_0)\),\\
      \, 3: & \quad \(t\sim{\rm Uniform}(\{1, 2, \ldots, T\})\)\\
      \, 4: & \quad \(\xi\sim\mathcal N(0,\bar\beta_{t})\)\\
      \, 5: & \quad \(\phi_t = \sqrt{\bar\alpha_t}\phi_0 + \xi\)\\
      \, 6: & \quad \(\phi_t^< = {\rm RG\ projection}(\phi_t,t)\)\\
      \, 7: & \quad \(\xi_\theta = {\rm DNN}_\theta(\phi_t^<,t)\)\\
      \, 8: & \quad \(\xi_\theta^< = {\rm RG\ projection}(\xi_\theta,t)\)\\ 
      \, 9: & \quad \(\xi^< = {\rm RG\ projection}(\xi,t)\)\\
      \, 10: & \quad \(g := \nabla_\theta \lambda_t ||\xi^< - \xi_\theta^<||^2\)\\
      \, 11: & \quad \(\theta := \theta - \alpha g\)\ \ \ (Update the model parameter.)\\
      \, 12: & \textbf{until} converged\\\hline 
    \end{tabular}
\end{table}

\noindent
Here, \({\rm DNN}_\theta\) denotes the denoising deep neural network without the RG projection, such as UNet. Also, the colored noise \(\xi\sim\mathcal N(0,\bar\beta_t)\) is sampled by using the DCT as:
\begin{table}[h]
    \centering
    \begin{tabular}{ll}\hline
      \multicolumn{2}{l}{\textbf{Algorithm 3}: Sampling of a colored noise \(\xi\sim\mathcal N(0,\bar\beta)\)\hspace*{106mm}}\\\hline
      \multicolumn{2}{l}{\, input: \((\bar\beta_k)_k\)}\\
      \, 1: & \textbf{for} all \(k\) \textbf{do}\\
      \, 2: & \quad \(\tilde\epsilon_k\sim \mathcal N(0,1)\)\\
      \, 3: & \quad \(\tilde\xi_k = \sqrt{\bar\beta_k}\tilde\epsilon_k\)\\
      \, 4: & \textbf{end for}\\
      \, 5: & \(\xi = {\rm DCT}^{-1}(\tilde\xi)\)\\
      \, 6: & \textbf{return} \(\xi\)\\\hline 
    \end{tabular}
\end{table}

After the training, we sample each data as follows (cf. Eqs.\eqref{Seq:bdiff}, \eqref{Seq:param:a}, and \eqref{Seq:param:b}, and Fig.~2c in the main text):
\begin{table}[H]
    \centering
    \begin{tabular}{rl}\hline
      \multicolumn{2}{l}{\textbf{Algorithm 4}: Sampling scheme of the RGDM\hspace*{118mm}}\\\hline
      \, 1: & \(\phi_T\sim\mathcal N(0,\bar\beta_T)\)\\
      \, 1: & \(\phi_T \leftarrow {\rm RG\ projection}(\phi_T, T)\)\\
      \, 1: & \textbf{for} \(t=T,\ldots,1\) \textbf{do}\\
      \, 2: & \quad \(\xi_\theta := {\rm DNN}_\theta(\phi_t,t)\)\\
      \, 2: & \quad \(\tilde\phi_t = {\rm DCT}(\phi_t)\),\ \ \(\tilde\xi_\theta = {\rm DCT}(\xi_\theta)\)\\
      \, 5: & \quad \textbf{for} \(|k|\leq c\Lambda_t\) \textbf{do}\\
      \, 5: & \qquad \(\tilde\epsilon_k\sim\mathcal N(0,1)\)\\
      \, 5: & \qquad \textbf{if} \(t=1\) \textbf{then} \(\tilde\epsilon_k := 0\)\\
      \, 5: & \qquad \(\displaystyle\tilde\phi_{t-1k} := \frac{1}{\sqrt{\alpha_{tk}}}\left(\tilde\phi_{tk}-\frac{\beta_{tk}}{\bar\beta_{tk}}\tilde\xi_{\theta k}\right) + \sqrt{\beta_{tk}}\tilde\epsilon_k\)\\
      \, 6: & \quad \textbf{end for}\\
      \, 6: & \quad \textbf{for} \(c\Lambda_t<|k|\leq c\Lambda_{t-1}\) \textbf{do}\\
      \, 6: & \qquad \(\tilde\phi_{t-1k} \sim \mathcal N(0,\bar\beta_{tk})\)\\
      \, 7: & \quad \textbf{end for}\\
      \, 9: & \quad \(\phi_{t-1} = {\rm DCT}^{-1}(\tilde\phi_{t-1})\)\\
      \, 12: & \textbf{end for}\\
      \, 12: & \textbf{return} \(\phi_{0}\) \\\hline 
    \end{tabular}
\end{table}

\noindent
Here, we note that \(\tilde\phi_{tk}\) with \(|k|>c\Lambda_t\) always satisfies \(\tilde\phi_{tk}=0\) during the sampling.

\subsubsection{\label{Sss:ns_param}Parameters of the noise schedule used in numerical experiments}
\begin{figure}[h]
    \centering
    \includegraphics[width=12cm]{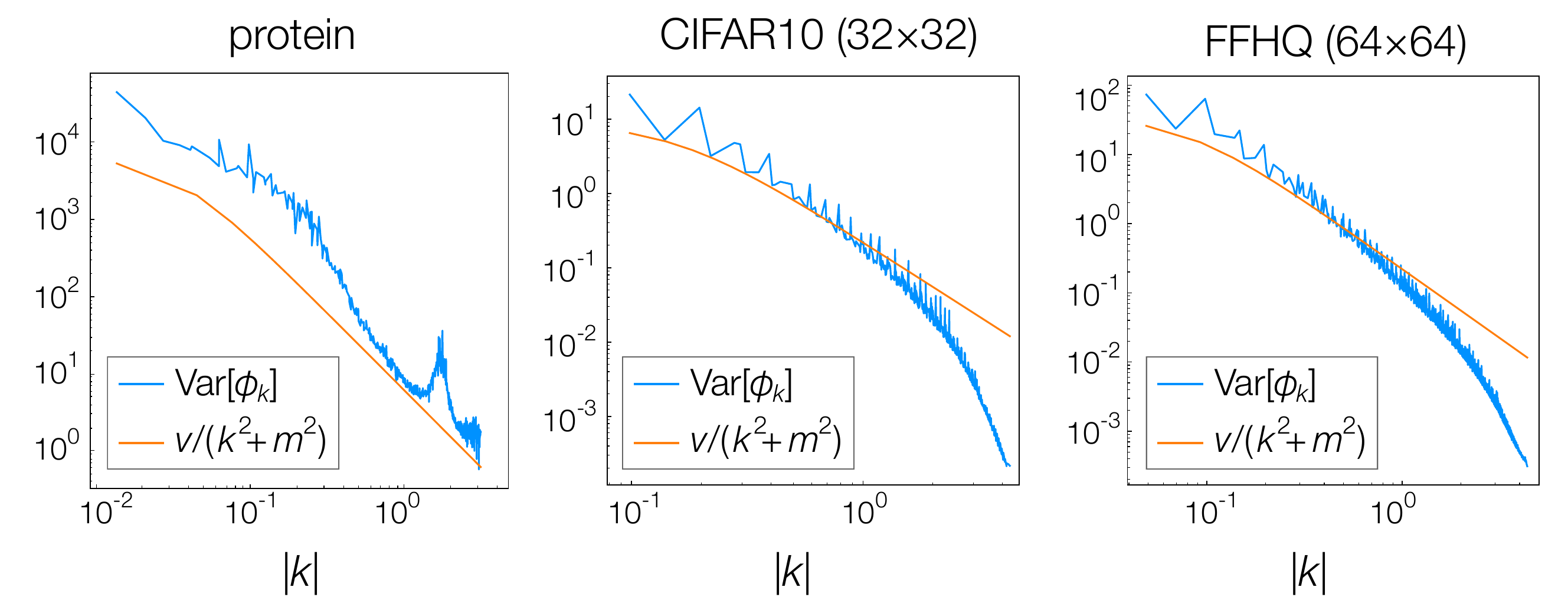}
    \caption{\label{Sfig:var_data}Plots of of the variance \({\rm Var}[\phi_k]\) and \(v/(k^2+m^2)\) on the protein data, CIFAR-10, and FFHQ datasets. The values of \(m\) and \(v\) in each dataset are summarized in Table~\ref{Stab:vm}}
\end{figure}

We here provide the hyperparameters in the noise schedules used in the numerical experiments.
As described above, we determine \(v\) and \(m\) in Eq.~\eqref{Seq:param:b} so that \(v/(k^2+m^2)\) roughly matches \({\rm Var}[\phi_k]\). 
In the protein structure prediction, we used protein data deposited in the PDB, whose carbon coordinate is written in the unit of angstrom. 
In  the image generation, we linearly rescaled each data so that the field \(\phi(x)\), which represents the RGB color in \(256\) levels of gradation, takes values within \([-1,1]\). 
We then used \(v\) and \(m\) in Table~\ref{Stab:vm}. 
Figure~\ref{Sfig:var_data} shows the behaviours of \({\rm Var}[\phi_k]\) and \(v/(k^2+m^2)\) in each dataset. 

\begin{table}[h]
    \centering
    \begin{tabular}{|c|c|c|}\hline
      dataset & \(v\) & \(m\) \\ \hline\hline
      protein & \((3.8)^2/2.3211\) & \(\pi/100\) \\ \hline
      CIFAR-10 (32\(\times\)32) & \((3\pi/20)^2\) & \(\pi/20\) \\ \hline
      FFHQ (64\(\times\)64) & \((3\pi/20)^2\) & \(\pi/40\) \\ \hline
    \end{tabular}
    \caption{\label{Stab:vm}Values of \(v\) and \(m\) used in the numerical experiments. These values are chosen so that \(v/(k^2+m^2)\) roughly matches the variance \({\rm Var}[\phi_k]\) in each dataset (cf. Eq.~\eqref{Seq:param:b} and Fig.~\ref{Sfig:var_data}).}
\end{table}

In choosing hyperparameters of the RGDM, the signal-to-noise ratio (SNR) \(v_{tk} = \bar\alpha_{tk}/\bar\beta_{tk}\) should satisfy the condition that \({\rm min}_kv_{1k}\) (\({\rm max}_k v_{Tk}\)) is sufficiently large (small) so that \(p_{t=1}\) and \(p_{t=T}\) approximately become \(p_{\rm data}\) and \(p_{\rm GS}\), respectively. We note that these choices ensure the conditons \(\Lambda_1\simeq\Lambda_{\rm UV}\) and \(\Lambda_T\ll \Lambda_{\rm IR}\) \cite{footnote1}. Below, we first summarize the noise schedule of the DDPM and then, explain how we determined the hyperparameters of the RGDM.

In the numerical experiments of the DDPM, we determined the noise schedule as in the reference~\cite{ho2020b}. Specifically, in DDPM with \(T=1000\), we used linearly increasing \(\beta_t\) with \(\beta_1=0.0001\) and \(\beta_{1000}=0.02\), which is the same choice as in the reference~\cite{ho2020b}. In DDPM with another \(T\), we used linearly increasing \(\beta_t\) and varying \(\beta_T\) so that the SNRs \(v_1\) and \(v_T\) are kept constant. We note that this also makes \(\bar\alpha_T\) and \(\bar\beta_T\) invariant when varying \(T\) since \(\bar\alpha_t\) and \(\bar\beta_t\) in the DDPM are related by \(\bar\alpha_t+\bar\beta_t=1\). 

\begin{figure}
    \centering
    \includegraphics[width=15cm]{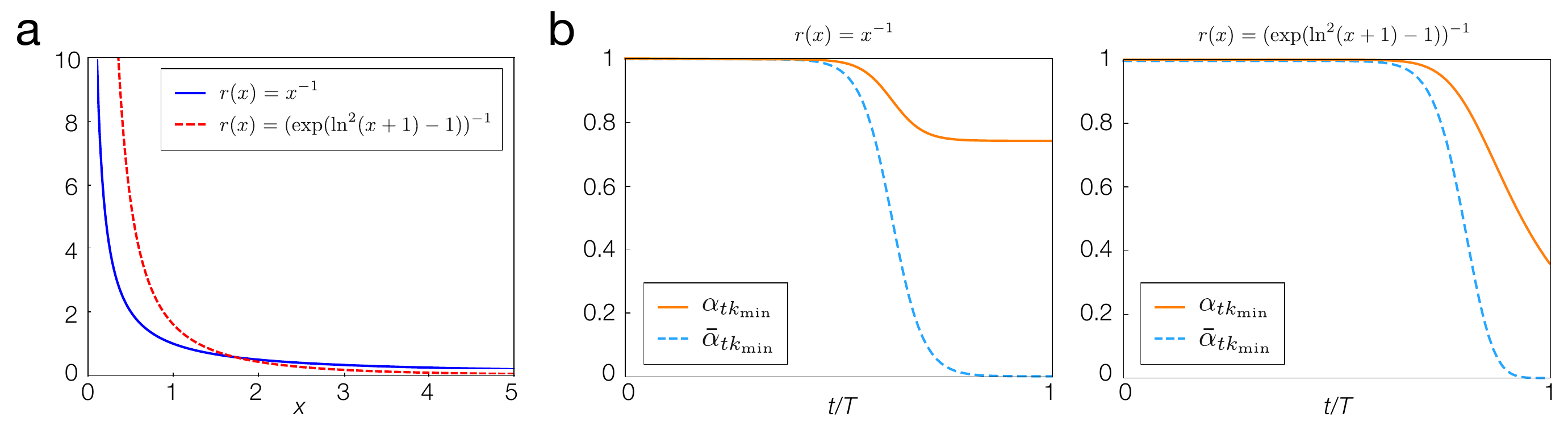}
    \caption{\label{Sfig:regulator}{\bf a,} Regulator functions used in image generation (\(r(x)=x^{-1}\)) and protein structure prediction (\(r(x)=(\exp(\ln(x+1)^2-1))^{-1}\)). {\bf b,} Typical behaviours of the noise schedules \(\alpha_{tk_{\rm min}}\) and \(\bar\alpha_{tk_{\rm min}}\) for each of the two regulators in panel {\bf a}. We note that \(\alpha_{tk}\) (\(\bar\alpha_{tk}\)) at different \(k\) is related to \(\alpha_{tk_{\rm min}}\) (\(\bar\alpha_{tk_{\rm min}}\)) as \(\alpha_{tk}=\alpha_{t+\tau\ln(k/k_{\rm min}),k_{\rm min}}\) (\(\bar\alpha_{tk}=\bar\alpha_{t+\tau\ln(k/k_{\rm min}),k_{\rm min}}\)). We set \(T=80\), and the other parameters are the same as in Table~\ref{Stab:Lam} in the left panel and Table~\ref{Stab:Lam_pro} in the right panel, respectively.}
\end{figure}

In image generation by the RGDM, we used a simple regulator \(r(x)=1/x\) (Fig.~\ref{Sfig:regulator}), which makes the SNR of each mode independent of wavenumber \(k\) as 
\begin{align}
    v_{tk} = \frac{\Lambda_t^2}{v}.
\end{align}
To make a fair comparison, in the numerical experiments, we used \(\Lambda_1\) and \(\Lambda_T\) in Table~\ref{Stab:Lam} so that the SNR at \(t=1,T\) becomes the same order of magnitude as the DDPM. 
We note that the values of \(\Lambda_1\) and \(\Lambda_T\) were fixed when varying \(T\) so that the SNRs at \(t=1,T\), i.e., \(v_{1k}\) and \(v_{Tk}\), are kept constant. Also, we used the cutoff wavenumber \(k_{\rm cutoff}\) for the RG projection layer (cf. Eq.~\eqref{Seq:k_cutoff}) as in Table~\ref{Stab:k_cutoff}. 

\begin{table}[h]
    \centering
    \begin{tabular}{|c|c|c|c|c|c|c|}\hline
      dataset & \(\Lambda_1\) & \(\Lambda_{T}\) & \(\Lambda_{\rm UV}\) & \(\Lambda_{\rm IR}\) & \({\rm min}_k v_{Tk}\) & \({\rm max}_k v_{Tk}\) \\ \hline\hline
      CIFAR-10 (32\(\times\)32) & 47.1215 & 0.00290 & \(\pi\) & \(\pi/32\) & 1e4 & 4e-5 \\ \hline
      FFHQ (64\(\times\)64) & 23.5608 & 0.00145 & \(\pi\) & \(\pi/64\) & 2.5e3 & 1e-5 \\ \hline
      DDPM & - & - & - & - & 1e4 & 4e-5 \\ \hline
    \end{tabular}
    \caption{\label{Stab:Lam} Values of \(\Lambda_1\) and \(\Lambda_T\) used in image generation (cf. Eq.~\eqref{Seq:param:Lam}). For the sake of completeness, we also show the minimum (maximum) value of the signal-to-noise ratio \(v_{tk}=\bar\alpha_{tk}/\bar\beta_{tk}\) at \(t=1\) \((t=T)\).}
\end{table}

\begin{table}[h]
    \centering
    \begin{tabular}{|c|c|}\hline
      dataset & \(k_{\rm cutoff}\) \\ \hline\hline
      CIFAR-10 (32\(\times\)32) & \(\pi/20\) \\ \hline
      FFHQ (64\(\times\)64) & \(\pi/40\) \\ \hline
    \end{tabular}
    \caption{\label{Stab:k_cutoff} Value of \(k_{\rm cutoff}\) used in image generation (cf. Eq.~\eqref{Seq:k_cutoff}).}
\end{table}

In protein structure prediction by the RGDM, we used a modified regulator \(r(x) = (\exp(\ln^2(x+1)-1))^{-1}\) (Fig.~\ref{Sfig:regulator}), which leads to wavenumber-dependent SNRs.
For this reason, one cannot make the SNRs of the RGDM and DDPM to be identical.
To make a fair comparison of these models, we thus consider the mutual information between \(\phi_0\) and \(\phi_t\) during the diffusion. As shown in Sec.~\ref{Ss:Disc_noise}, in the RGDM, the mutual information between \(\phi_0\sim \prod_k \mathcal N(\phi_{0k};0,vk^{-2})\) and \(\phi_t\) becomes \(I_{\rm RGDM}(\phi_0:\phi_t) = \sum_k (1/2)\ln(1+r_{tk})\). Similarly, in the DDPM, the mutual information between \(\phi_0 \sim \prod_k \mathcal N(\phi_{0k};0,1)\) and \(\phi_t\) becomes \(I_{\rm DDPM}(\phi_0:\phi_t) = \sum_k (1/2)\ln(1/\bar\beta_{tk})\). By comparing \(I_{\rm RGDM}\) and \(I_{\rm DDPM}\), we determined \(\Lambda_0\) and \(\tau\) in the RGDM by the following conditions: 
\begin{align}
   \min_k (1+r_{t=1,k}) &= 1 + r_{t=1,k_{\rm max}} = 1/\bar\beta_{{\rm DDPM},t=1},\\
   \max_k (1+r_{t=T,k}) &= 1+ r_{t=T,k_{\rm min}} = 1/\bar\beta_{{\rm DDPM}, t=T}.
\end{align}
Since we choose \(T\) as a function of protein length \(N\) as \(T(N) = T(N_0) + \tau\ln(N/N_0)\) (see Methods in the main text), \(\Lambda_1=\Lambda_0e^{-1/\tau}\) and \(\tau\) are independent of  protein length as summarized in Table~\ref{Stab:Lam_pro}. Also, we took the cutoff wavenumber \(k_{\rm cutoff}\) for the RG projection layer (cf. Eq.~\eqref{Seq:k_cutoff}) as \(k_{\rm cutoff} = k_{\rm min}\) to discard \(k\)-th field components with \(r_{tk}<\max_k r_{Tk}\), which are already integrated out as described in the main text. In addition, in protein structure prediction, we also discarded the Fourier components of the predicted noise \(\xi_\theta(\phi_t,t)\) with \(r_{tk}>\min_k r_{1k}\) since \(\phi_{tk}\) can be regarded as already denoised in the generation process. Specifically, the cost function during the training (Eq.~\eqref{Seq:cost}) is modified as 
\begin{align}
    L_\theta = \sum_{t=1}^T \lambda_t\ \mathbb E_{\phi_0\sim p_0(\phi),\xi_t\sim\mathcal{N}(0,\bar\beta_t)}
    \left[\sum_{k; \max_k r_{Tk}\leq r_{tk}\leq \min_k r_{1k}} ||\xi_{\theta k}(\phi_t,t)-\xi_{tk}||^2\right],
\end{align}
where we also modified \(\lambda_t\) as 
\begin{align}
    \lambda_t \propto \left(\sum_{k; \max_k r_{Tk}\leq r_{tk}\leq \min_k r_{1k}} 1 \right)^{-1} 
    = \left( \# \text{ of } k \text{ s.t. } \max_k r_{Tk}\leq r_{tk}\leq \min_k r_{1k}\right)^{-1}.
\end{align}

\begin{table}[h]
    \centering
    \begin{tabular}{|c|c|c|c|c|c|c|}\hline
      dataset & \(\Lambda_1\) & \(\tau\) & \(\Lambda_{\rm UV}\) & \(\Lambda_{\rm IR}\) \\ \hline\hline
      protein & 27.7 & 8.17 & \(\pi\) & \(\pi/N\) \\ \hline
    \end{tabular}
    \caption{\label{Stab:Lam_pro} Values of \(\Lambda_1\) and \(\tau\) used in protein structure prediction (cf. Eq.~\eqref{Seq:param:Lam}). We note that \(\Lambda_1\) and \(\tau\) are kept constant when varying protein length \(N\). Instead, the number of generation steps \(T\) are varied as \(T(N) = T(N_0) + \tau\ln(N/N_0)\) with \(T(32)=80\).}
\end{table}

\subsubsection{\label{Sss:nn_param}Neural networks in numerical experiments}
\begin{figure}
    \centering
    \includegraphics[width=17cm]{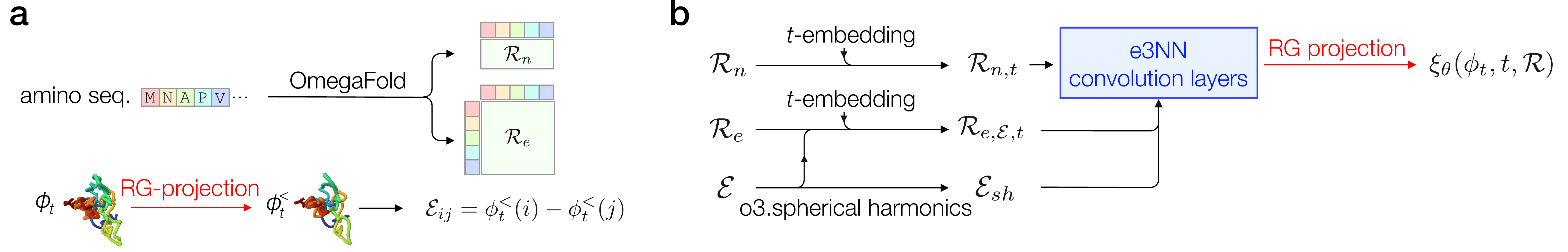}
    \caption{\label{Sfig:e3NN} Structure of deep neural network to predict the colored noise added on \(\phi_t\) by the RG forward diffusion in the protein structure prediction. {\bf a,} We first create feature tensors (node embedding \(\mathcal R_e\) and edge embedding \(\mathcal R_n\)) by using OmegaFold \cite{wu2022}. Also, we prepare edge tensor \(\mathcal E\) of \(\phi_t\) as \(\mathcal E_{ij}=\phi_t^<(i)-\phi_t^<(j)\) with \(1\leq i,j\leq N\). {\bf b,} After embedding the information about \(t\) and mapping \(\mathcal E\) to the spherical harmonics representation \(\mathcal E_{sh}\) by using the o3.spherical harmonics library in Python, we apply e3NN convolution layers and predict the colored noise \(\xi_\theta(\phi_t,t,\mathcal R)\).}
\end{figure}

We here provide technical details of the denoising deep neural networks (DNNs) used in the numerical experiments. We note that the numerical codes are available on GitHub: \url{https://github.com/kantamasuki/RGDM}. 
As mentioned in the Method, we used denoising DNNs based on e3NN and UNet in the application for the protein structure prediction and image generation, respectively. 
In the protein structure prediction, following the previous study~\cite{jing2023}, we first create the feature tensors of each amino sequence \(\mathcal R\), namely, the node embedding \(\mathcal R_{n}\) and the edge embedding \(\mathcal R_e\), by using OmegaFold \cite{wu2022}. We note that the edge and node embeddings basically extract the positional information about the \(i\)-th amino node (\(1\leq i\leq N\)) and the \((i,j)\)-th amino edge (\(1\leq i,j\leq N\)) of a protein chain of length \(N\) (Fig.~\ref{Sfig:e3NN}a). Also, we create the edge tensor of \(\phi_t^<\), the lower wavenumber components of \(\phi_t\), as \(\mathcal E_{ij}=\phi_t^<(i)-\phi_t^<(j)\) with \(1\leq i,j\leq N\). After embedding the information about \(t\), we apply e3NN convolution layers to these tensors and predict \(\xi_\theta(\phi_t,t,\mathcal R)\), i.e., the colored noise added by the forward RG diffusion (Fig.~\ref{Sfig:e3NN}b). We note that the e3NN networks are equivariant under the \(O(3)\)-rotation and translation. During the training, we took the number of warmup steps as \(10^4\) and updated the model parameters by the Adam optimizer with a learning rate \(3\times 10^{-4}\). Also we took the batch size \(1\) since the tensor size of \(\phi_t\) depends on the length of each protein \(N\).

\begin{figure}
    \centering
    \includegraphics[width=10cm]{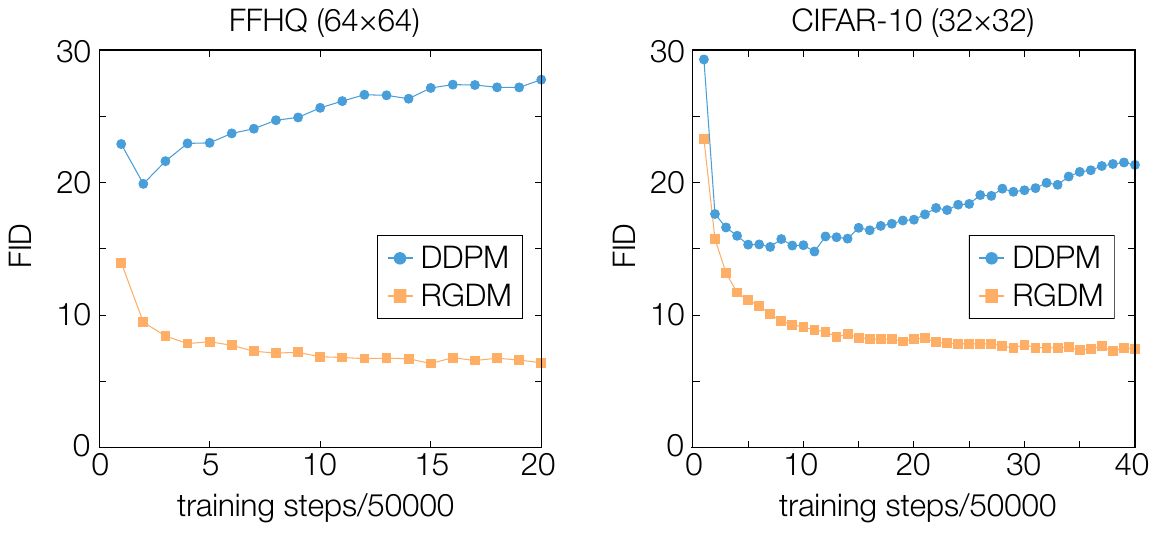}
    \caption{\label{Sfig:fid_tra} Sampling quality of the RGDM and DDPM plotted against the number of training steps, where the models are unconditionally trained on the FFHQ dataset (left panel) and the CIFAR-10 dataset (right panel) with total generation steps \(T=300\). The sampling quality at each datapoint is obtained by evaluating the frech\'et interception distance (FID) \cite{heusel2017a} with 1\(\times 10^4\) generated samples.}
\end{figure}

In the image generation, we used the UNet structure as the denoising DNN, which is the same as in the DDPM proposed by Ho et al \cite{ho2020b}. In our UNet architecture, we set the channel number as \(128\) and the channel multiplier as \([1,2,2,2]\). For image data with resolution \(L\times L\), attention layers were applied at resolution \(L/2\times L/2\). These hyperparameters are summarized in Table~\ref{Stab:Unet}. All the models were optimized with Adam optimizer with a learning rate \(2\times 10^{-4}\), a dropout ratio 0.1, and warmup steps 5000. Also, we took the batch size 128. The weight of model parameters was calculated with exponential moving average of rate \(0.9999\). 
During the numerical experiments of image generation, we find that the model performance of DDPM saturates at finite training steps, while the performance of the RGDM steadily improves (Fig.~\ref{Sfig:fid_tra}). 
We argue that this stability of the model performance results from both the noise schedule and the RG projection schemes of the RGDM. A similar behaviour of the model performance was observed during the training of the RGDM without the RG projection layers, but the addition of the RG projection layers further stabilized the training and resulted in a better performance as presented in the main text.

\begin{table}[h]
    \centering
    \begin{tabular}{|l|c|c|c|c|c|}\hline
      Experiment & Channels & Attention res. & ResBlocks/stage & Channel multiplier & Dropout \\ \hline\hline
      CIFAR-10(32\(\times 32\)) & 128 & 16 & 2 & [1,2,2,2] & 0.1 \\ \hline
      FFHQ (64\(\times\)64) & 128 & 32 & 2 & [1,2,2,2] & 0.1 \\ \hline
    \end{tabular}
    \caption{\label{Stab:Unet} UNet architectures used in the numerical experiments of image generation.}
\end{table}

\subsubsection{\label{Sss:sampled_data}Examples of sampled data in numerical experiments}
\begin{figure}
    \centering
    \includegraphics[width=15cm]{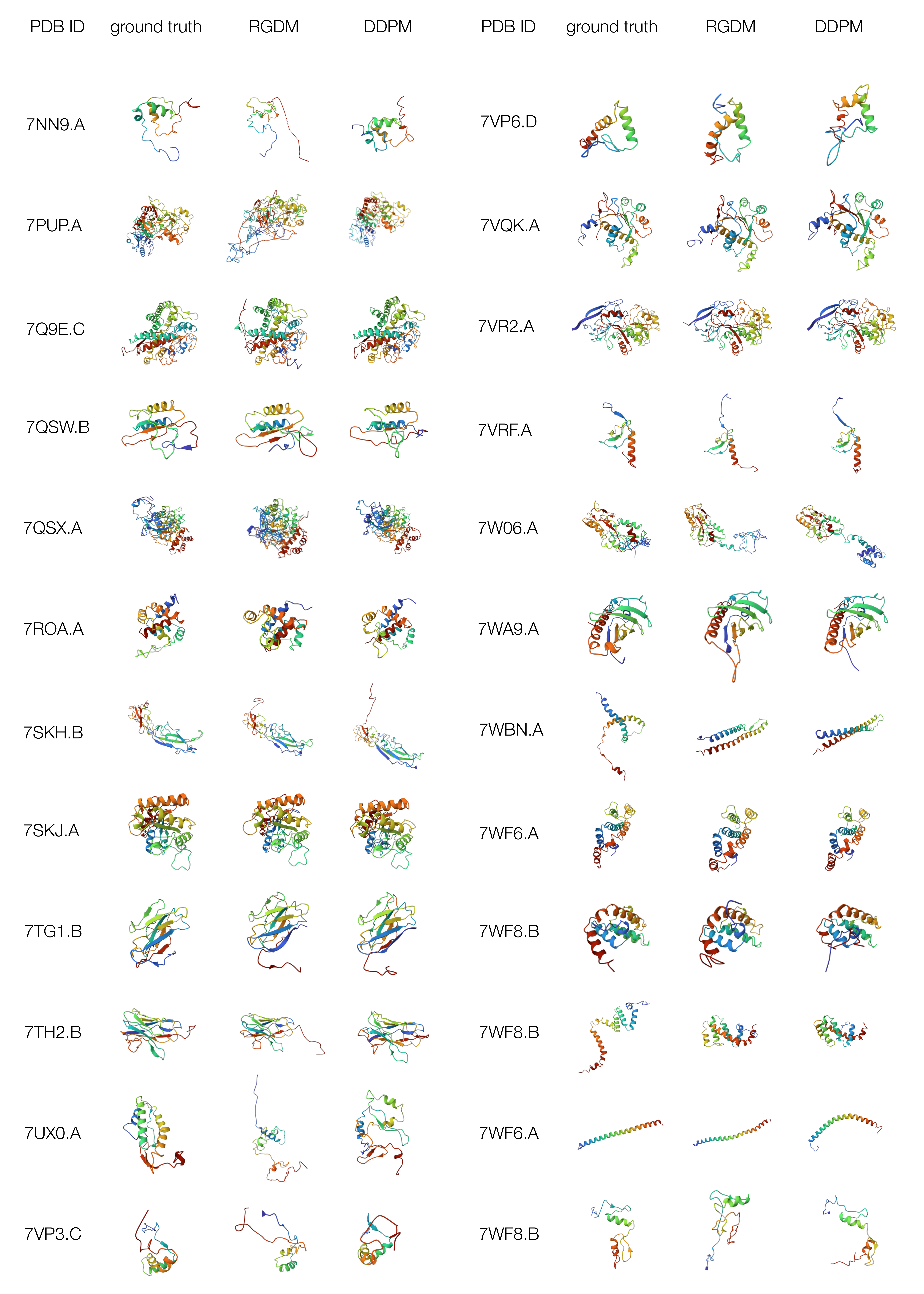}
    \caption{\label{Sfig:pro_samp}Protein structures generated by the RGDM and DDPM in the numerical experiments.}
\end{figure}

\begin{figure}
    \centering
    \includegraphics[width=16cm]{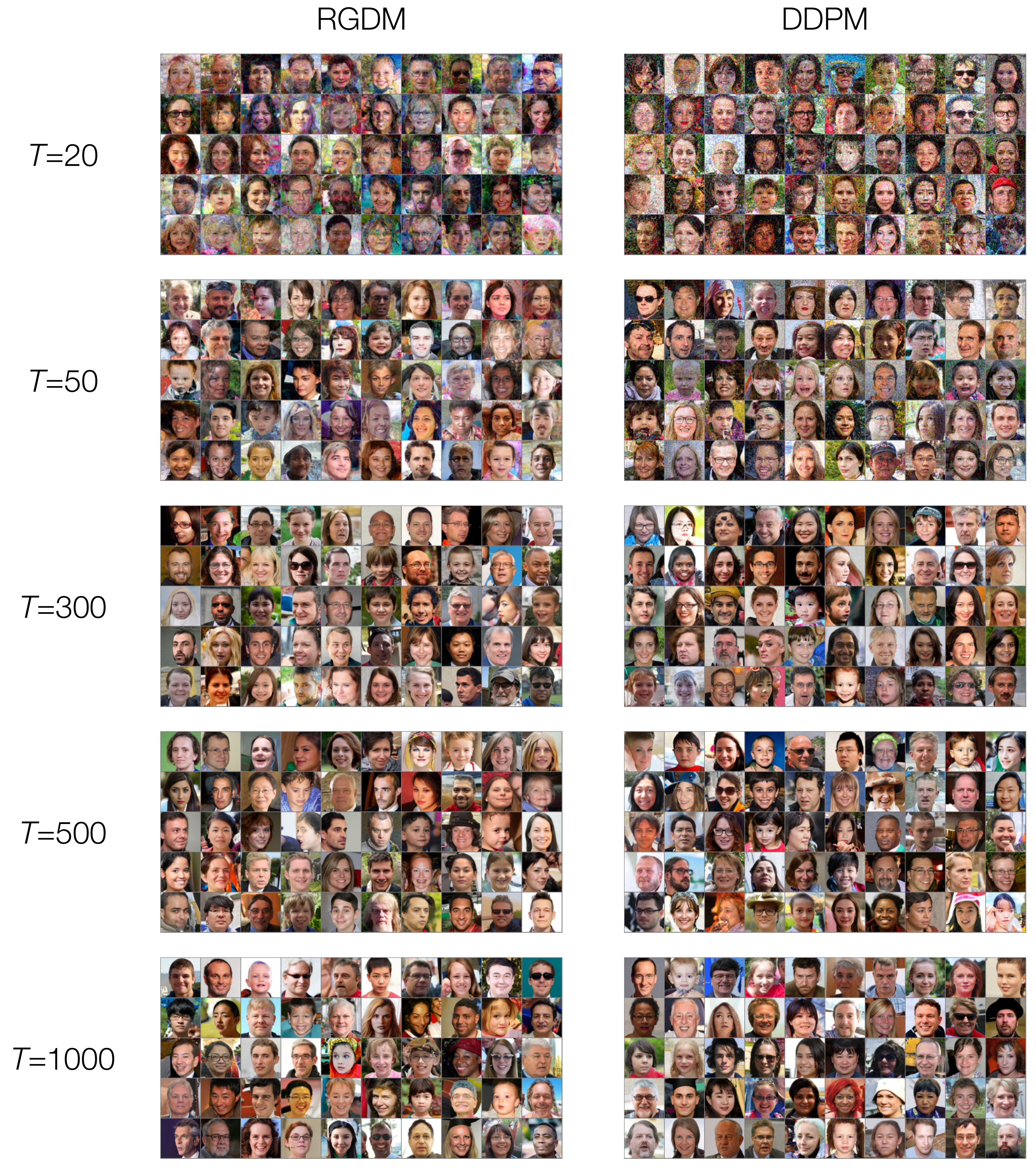}
    \caption{\label{Sfig:ffhq_samp}Images generated by the the RGDM and DDPM trained on the FFHQ dataset with image resolution \(64\times 64\) at the total number of generating steps \(T=20, 50, 300, 500, 1000\).}
\end{figure}

\begin{figure}
    \centering
    \includegraphics[width=16cm]{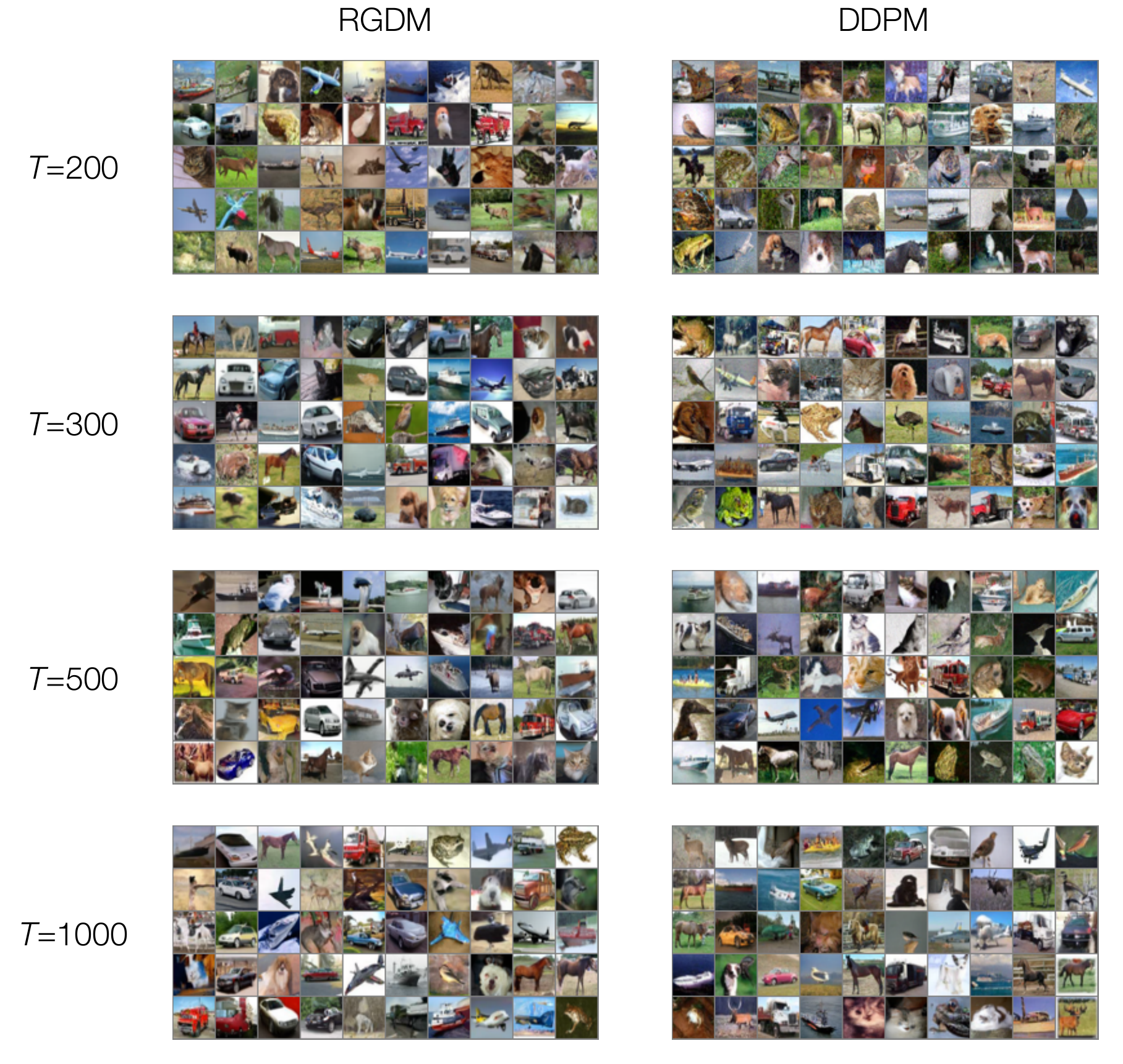}
    \caption{\label{Sfig:cifar_samp}Images generated by the the RGDM and DDPM trained on the CIFAR-10 dataset with image resolution \(32\times 32\) at the total number of generating steps \(T=200, 300, 500, 1000\).}
\end{figure}

We here provide the samples generated by the RGDM and DDPM in the numerical experiments.
Figure~\ref{Sfig:pro_samp} shows the protein structures predicted by the diffusion models, and Figures~\ref{Sfig:ffhq_samp} and \ref{Sfig:cifar_samp} show the typical samples of image data, respectively.

\clearpage

\subsection{\label{Ss:Disc_noise}Discussions about the noise schedule of the renormalization group diffusion model}
In the studies of diffusion model, it is known that all the noise schedules of the wavenumber-independent (i.e., white) noise are mathematically equivalent in the continuous time limit in the sense that they can be rescaled to each other \cite{kingma2010,yang2023a}. Here, in a similar manner, we first show that the noise schedule of the RGDM can cover all the wavenumber-dependent (i.e., colored) noise schedules by appropriately modifying the RG cutoff function \(K_\Lambda(k)\). We then discuss a condition that naturally characterizes the noise schedule and the choice of \(K_\Lambda(k)\) used in the RGDM.

We consider a class of diffusion models with wavenumber-dependent noise schedules (\(\bar\alpha_{tk}, \bar\beta_{tk}\)). In such models, the distribution of \(\phi_t\) is given by
\begin{align}
    p_t(\phi_t)
    &= \int [d\phi_0]\, \prod_k \mathcal{N}(\phi_{tk};\sqrt{\bar\alpha_{tk}}\phi_{0k},\bar\beta_{tk}) p_0(\phi_0)\\
    &= \int [d\phi_0]\prod_k \frac{1}{\sqrt{\bar\alpha_{tk}}}\ \mathcal{N}\left(\frac{\phi_{tk}}{\sqrt{\bar\alpha_{tk}}}; \phi_{0k}, \frac{1}{v_{tk}}\right) p_0(\phi_0)\\
    &= c_t\, p^*_{v_t}\left(\frac{\phi_t}{\sqrt{\bar\alpha_t}}\right),\label{Seq:pt_pvt}
\end{align}
where \(c_t = \prod_k 1/\sqrt{\bar\alpha_{tk}}\) is the normalization constant, \(v_{tk} = \bar\alpha_{tk}/\bar\beta_{tk}\) is the signal-to-noise ratio (SNR) of the \(k\)-th mode at \(t\), and we define \(p^*_v(\phi)\) for \(v=(v_k)_k\) by 
\begin{align}
    p^*_{v_t}(\phi) &= \int [d\phi_0] \prod_k \mathcal{N}(\phi_k;\phi_{0k}, v_{k}^{-1})p_0(\phi_0).\label{Seq:pv}
\end{align} 
Notably, Eq.\eqref{Seq:pt_pvt} implies that the intermediate distribution in the diffusion model, \(p_t(\phi_t)\), is characterized only by the \(k\)-dependent SNR \(v_t = (v_{tk})_k\) up to the rescaling of the field. In fact, the optimal value of the denoising function \(\xi_\theta(\phi_t,t)\), which minimizes the cost function \(L_\theta\) in Eq. (10) in the Method, can also be characterized by \(v_t\) as 
\begin{align}
    \xi_\theta(\phi_t,t) = \sqrt{\alpha_{t}} \xi^*_{v_t}\left(\frac{\phi_t}{\sqrt{\alpha_t}},t\right),
\end{align}
where \(\xi_v^*\) minimizes
\begin{align}
    L &= \mathbb E_{\phi_0\sim p_0(\phi_0), \xi\sim\mathcal N(0,v_t^{-1})} 
    \left[
        || \xi_{v_t}^*(\phi_0+\xi,t) - \xi ||^2
    \right].
\end{align}
Therefore, two different diffusion models become ``equivalent'' up to the rescaling of the field if the trajectories of \(v_t\) (i.e., \(\mathcal C_v\equiv\{v_t;t\in[0,T]\}\)) coincide.
In this sense, one can show that the RGDM covers all the diffusion models with wavenumber-dependent noise schedules. To see this, we recall that the SNRs of the RGDM is given by (cf. Eq. (9) in the main text) 
\begin{align}
    v_{tk} = k^2\frac{K_{\Lambda_t}(k)}{1-K_{\Lambda_t}(k)}.
\end{align}
Therefore, one can always construct an RGDM that is equivalent to the diffusion model with an arbitrary SNR trajectory \(\mathcal C_v\) by choosing the RG cutoff function \(K_{\Lambda_t}(k)\) as \(K_{\Lambda_t}(k)\!=\!v_{tk}/(v_{tk}+k^2)\).

On the other hand, however, mathematically equivalent diffusion models upon the rescaling do not necessarily lead to equivalent performances in practice. For example, although all diffusion models with white noise become equivalent up to rescaling in the continuous time limit, they show completely different performances depending on a specific choice of noise schedules \cite{karras2022a,chen2023,hoogeboom2023}. Therefore, it is meaningful to discuss which choice will be an optimal one for the stable and efficient probability flow in the RGDM among the mathematically equivalent choices. 

We here argue that an optimal schedule of the wavenumber-dependent noise can be characterized by the following conditions, which lead to the noise schedule obtained in the RGDM:
\begin{enumerate}
    \item[(A)] The Gaussian distribution \(p_{\rm GS}(\phi)\propto e^{-\int_x(\nabla\phi)^2/2}\) loses its information along the forward diffusion process as a function of \(|k|/\Lambda_t\).
    \item[(B)] The Gaussian distribution \(p_{\rm GS}(\phi)\propto e^{-\int_x(\nabla\phi)^2/2}\) is the steady-state solution of the forward diffusion process.
\end{enumerate}
To show that these conditions indeed lead to the noise schedule of the RGDM, we first note that the information about \(\phi_0\) included in \(\phi_t\) will be diminished along the forward diffusion \(p(\phi_t|\phi_0) = \prod_k \mathcal N(\phi_{tk};\sqrt{\bar\alpha_{tk}}\phi_{0k},\bar\beta_{tk})\) as
\begin{align}
    I(\phi_0:\phi_t) &= \int [d\phi_0][d\phi_t] p(\phi_0,\phi_t)\ln \frac{p(\phi_0,\phi_t)}{p_0(\phi_0)p_t(\phi_t)}\\
    &= \sum_k \frac{1}{2} \ln\left(1+\frac{v_{tk}}{k^2}\right),
\end{align}
where \(I(\phi_0:\phi_t)\) is the mutual information between \(\phi_0\) and \(\phi_t\), and we use \(p(\phi_0,\phi_t) = p(\phi_t|\phi_0)p_{\rm GS}(\phi_0)\). 
With this expression, the condition (A) above can be simply rephrased as the statement that \(v_{tk}/k^2\) is a function of \(|k|/\Lambda\) and monotonically decreasing as a function of $t$. More specifically, this ensures that the SNR can be written as \(v_{tk}=k^2r(k^2/\Lambda_t^2)\) with a function monotonically decreasing function \(r(x)\).
We note that this is precisely what has been achieved in the RGDM, where $r(x)$ is nothing but a regulator. 
Meanwhile, the condition (B) imposes the constraint on \(\bar\alpha_{tk}\) and \(\bar\beta_{tk}\) such that \(\bar\beta_{tk}=k^{-2}(1-\bar\alpha_{tk})\). 
Combining these two conditions, we thus arrive at the expressions,
\begin{align}
    \bar\alpha_{tk} &= \frac{r(k^2/\Lambda_t^2)}{1+r(k^2/\Lambda_t^2)},\ \bar\beta_{tk} = \frac{1}{k^2}\frac{1}{1+r(k^2/\Lambda_t^2)},
\end{align}
which give the noise schedule of the RGDM (see Eqs.~(5) and (9) in the main text). 
We expect that the conditions (A) and (B), which reflect the spirit of RG procedures in the above sense, might serve as a guiding principle for noise scheduling of the diffusion models with the colored noise.


\begin{thebibliography}{30}
\bibitem[64]{footnote1} Strictly speaking, since the cutoff function \(K_\Lambda(k)\) smoothly turns from one to zero ``around'' \(|k|=\Lambda\), we actually take \(\Lambda_{t=0}\) larger than \(\Lambda_{\rm UV}\) so that \(K_{\Lambda_{t=0}}(k)\simeq1\) is satisfied for all \(k\).
\end{thebibliography}
\end{document}